\documentclass[a4paper,fleqn,usenatbib]{mnras}
\usepackage{newtxtext,newtxmath}
\usepackage[dvipsnames]{xcolor}
\usepackage{ae,aecompl}
\usepackage{graphicx}	
\usepackage{amsmath}	
\usepackage{mathtools}
\usepackage{courier}

\usepackage{hyperref}
\hypersetup{draft=false}

\title[DEVILS: The sSFR-M$_{\star}$ plane part II]{Deep Extragalactic VIsible Legacy Survey (DEVILS):  The sSFR-M$_{\star}$ plane part II: Starbursts, SFHs and AGN Feedback}
\author[L. J. M. Davies]{L. J. M. Davies$^{1}$\thanks{E-mail:
 luke.j.davies@uwa.edu.au},  J. E. Thorne$^{1}$, S. Bellstedt$^{1}$, R. H. W. Cook$^{1}$, M. Bravo$^{2}$,  A. S. G. Robotham$^{1,3}$, \newauthor C. del P. Lagos$^{1,3,4}$, S. Phillipps$^{5}$,  M. Siudek$^{6,7}$, B. W. Holwerda$^{8}$, M. N. Bremer$^{5}$, J. D'Silva$^{1}$, \newauthor and S. P. Driver$^{1}$ \\
 $^{1}$ ICRAR, The University of Western Australia, 35 Stirling Highway, Crawley, WA 6009, Australia \\
$^{2}$ Department of Physics \& Astronomy, McMaster University, 1280 Main Street W, Hamilton, ON, L8S 4M1, Canada \\
$^{3}$ ARC Centre of Excellence for All Sky Astrophysics in 3 Dimensions (ASTRO 3D) \\
$^{4}$ Cosmic Dawn Center (DAWN), Denmark \\
$^{5}$ Astrophysics Group, School of Physics, University of Bristol, Bristol BS8 1TL, UK \\
$^{6}$ Institute of Space Sciences (ICE, CSIC), Campus UAB, Carrer de Can Magrans, s/n, 08193 Barcelona, Spain \\
$^{7}$Institut de Física d’Altes Energies (IFAE), The Barcelona Institute of Science and Technology, 08193 Bellaterra (Barcelona), Spain\\
$^{8}$ Physics \& Astronomy Department, University of Louisville, Louisville, KY 40292, USA \\
}

\date{Accepted XXX. Received YYY; in original form ZZZ}

\pubyear{2025}

\begin{document}
\label{firstpage}
\pagerange{\pageref{firstpage}--\pageref{lastpage}}
\maketitle

\begin{abstract}
In part I of this series we discussed the variation of star-formation histories (SFHs) across the specific star formation rate - stellar mass plane (sSFR-M$_{\star}$) using the Deep Extragalactic VIsible Legacy Survey (DEVILS).  Here we explore the physical mechanisms that are likely driving these observational trends, by comparing the properties of galaxies with common recent SFH shapes. Overall, we find that the processes shaping the movement of galaxies through the sSFR-M$_{\star}$ plane can be be largely split into two stellar mass regimes, bounded by the minimum SFR dispersion ($\sigma_{SFR}$) point.  At lower stellar masses we find that large $\sigma_{SFR}$ values are likely observed due to a combination of stochastic star-formation processes and a large variety in absolute sSFR values, but relatively constant/flat SFHs. While at higher stellar masses we see strong observational evidence that Active Galactic Nuclei (AGN) are associated with  rapidly declining SFHs, and that these galaxies reside in the high $\sigma_{SFR}$ region of the plane. As such, we suggest that AGN feedback, leading to galaxy quenching, is the primary driver of the high $\sigma_{SFR}$ values. These results are consistent with previous theoretical interpretations of the $\sigma_{SFR}$-M$_{\star}$ relation.   
             
\end{abstract}


\section{Introduction}

The relationship between a galaxy's stellar mass and specific star-formation rate, the sSFR-M$_{\star}$ plane, is a key diagnostic tool in understanding the processes that shape the evolution of galaxies \citep[$e.g.$][]{Elbaz07,Noeske07,Salim07,Whitaker12, Johnston15, Davies16b}. In this plane the bulk of star-forming galaxies lie on a tight linear sequence, known as the star-forming galaxy `main' sequence (SFS), with more massive galaxies having higher star-formation rates. This relation is largely thought to arise from the self-regulation of star-formation \citep[$i.e.$][]{Bouche10, Daddi10, Genzel10, Lagos11, Lilly13, Dave13, Mitchell16}, where the inflow of gas for star-formation is balanced by the formation of new stars and outflows driven by feedback events ($i.e.$ from Supernovae, SNe, and AGN). However, the sSFR-M$_{\star}$ plane does not only contain the SFS. The plane is populated with other galaxies that likely once resided on the SFS but over the course of their lifetime have evolved away from the sequence, as they left a self-regulated state \citep[$e.g.$][]{Tacchella16}. Identifying these populations and understanding the root causes that have driven them off the SFS is paramount to understanding the galaxy evolution process. The main trend of the evolution of galaxies off the SFS, is from the self-regulated star-forming state, with ready access to gas reservoirs, to a passive quiescence state where they no longer turn gas into stars. This transition is witnessed as the increasing number of quiescent systems as the Universe evolves \citep[$e.g.$][]{Damen09} and the overall decline in cosmic star-formation since $z\sim2$ \citep[e.g.][]{Bellstedt20}.    

Many parameters are known to affect a galaxy's star-formation rate and therefore stellar mass growth, potentially leading to these quenching events, namely Active Galactic Nuclei (AGN) feedback \citep{Kauffmann04,Fabian12}, stellar feedback \citep{Dekel86, DallaVecchia08, Scannapieco08}, mergers \citep[$e.g.$][]{Bundy04, Baugh06, Kartaltepe07, Bundy09,Jogee09,deRavel09,Lotz11,Robotham14}, morphological/structure evolution \citep{Conselice14, Eales15}, gas fuelling \citep{Kauffmann06,Sancisi08, Mitchell16}, secular quenching \citep{Schawinski14, Barro13}, environmental quenching \citep[$e.g.$][]{Giovanelli85, Peng10, Cortese11, Darvish16, Davies19b}, etc. Exploring the relative contribution of these processes to shaping the populations in the sSFR-M$_{\star}$ plane, and the fine details of $how$ each of the processes occur, will help us better constrain the evolution of galaxies and their transition from self-regulated star-forming galaxies, to passive dead systems.   

Previous studies have typically used the position of a galaxy with respect to the SFS to select sub-populations of galaxies which have either transitioned off the SFS at some point in the past (essentially systems with SFRs far below the SFS - deemed to be quiescent), are currently going through a `star-burst' event  (systems which sit significantly above the SFS) and galaxies that have recently left the SFS and may be actively quenching  \citep[systems that sit just below the SFS, in what would typically be deemed the `green valley', $e.g.$][]{Elbaz11, Schawinski14,Bremer18,Phillipps19, Salim23}. However, this picture is a complicated one. The processes that cause galaxies to leave the locus of the SFS likely vary as a function of stellar mass, morphology, structure, environment, available gas supply and the presence of an AGN. In addition, the SFS itself is not a static entity and evolves with time, likely in a manner that is not self-similar across stellar masses and environments \citep[$e.g.$][]{Lee15, Leslie20, Thorne21}. As such, even defining a galaxy's position relative to the SFS, or what that means in terms of its current star formation, is fraught with difficulty; let alone picking apart the dominant astrophysical mechanisms that caused an individual galaxy to leave this relation. Consequently, this has been a very active field of research over the last decade or more.  

What may be most important in studying the evolution of galaxies using the SFR-M$_{\star}$ plane is not the position of the galaxy with respect to the SFS, but \textit{how} the galaxy is moving \textit{through} the sSFR-M$_{\star}$ plane \citep[$e.g.$][]{Sanchez18, Iyer18, Ciesla17, Ciesla21, Arango24}. This essentially tells us the current trajectory of a galaxy as it either evolves on the SFS, or leaves the SFS via a quenching/star-burst event. To parameterise this, one must look to recent star formation histories \citep[SFH,][]{Madau98, Kauffmann03, Bellstedt20}, for which we can derive the past SFR and stellar mass of a particular galaxy using its current distribution of stellar ages. The most robust SFHs are derived through high signal-to-noise spectroscopic observations \citep[$e.g.$][]{Fernandes05}. However, these are expensive, requiring significant investment of telescope time and therefore can only be derived for a relatively small number of sources. 

In our recent studies with both the Galaxy and Mass Assembly Survey \citep[GAMA, $e.g.$][]{Bellstedt20} and the Deep Extragalactic VIsible Legacy Survey \citep[DEVILS, $e.g.$][]{Thorne21,Thorne22} we instead used the \textsc{ProSpect} spectral energy fitting code \citep{Robotham20} to derive the recent SFHs of galaxies using photometric data and spectroscopic redshifts alone. This process has been shown to be robust in measuring SFHs for galaxies (see discussion in Section \ref{sec:validity}), but can easily be applied to large datasets, such as the aforementioned surveys.    

In the first paper in this series (Davies et al 2025, hereafter paper I), we explored the variation in \textsc{ProSpect}-derived SFHs across the sSFR-M$_{\star}$ plane and their evolution with time. We selected galaxies using position relative to the SFS (in a similar approach to previous studies), and showed that while these selections identified galaxies with common SFHs, there is a large spread in SFHs at a given position in the sSFR-M$_{\star}$ plane. Most notably, the majority of high stellar mass galaxies that sit on the SFS, actually show rapidly declining SFHs. These systems would traditionally have been classed as having constant self-regulated star-formation based on their position in the plane, but in-fact appear to be undergoing a quenching event - highlighting the difficulty in selecting populations based on position in the sSFR-M$_{\star}$ plane alone. We then identified new regions in the sSFR-M$_{\star}$ plane that have common recent SFHs (described in the following section), and explore how these populations evolve with time. Finally, we showed how the combination of these different populations evolving through the sSFR-M$_{\star}$ plane leads to the overall evolution of the SFS. For further details we refer the reader to paper I. 

In this paper we expand upon this to explore commonalities in the physical properties of galaxies with similar recent SFHs. Specific emphasis is placed on galaxies that are significantly evolving through the sSFR-M$_{\star}$ plane, with the aim of identifying the astrophysical drivers that lead to their movement off the SFS. In this paper we aim to provide a comprehensive description of the possible astrophysical mechanisms which cause galaxies to move through the sSFR-M$_{\star}$ plane. For the casual reader, we refer them to Figure \ref{fig:cartoon} which aims to provide an overall summary of the trends/results found in paper I and II of this series.

\section{Data and Sample Selection}

The bulk of the data description for this work is presented in paper I in this series. However, for completeness we briefly summarise here.

\subsection{The Deep Extragalactic VIsible Legacy Survey}

DEVILS is a spectroscopic survey undertaken at the Anglo-Australian Telescope (AAT), which aimed to build a high completeness ($>$85\%) sample of $\sim$50,000 galaxies to Y$<$21\,mag in three well-studied deep extragalactic fields: D10 (COSMOS), D02 (ECDFS) and D03 (XMM-LSS). The survey will provide the first high completeness sample at $0.3<z<1.0$, allowing for the robust parametrisation of group and pair environments in the distant Universe. The science goals of the project are varied, from the environmental impact on galaxy evolution at intermediate redshift, to the evolution of the halo mass function over the last $\sim$7\,billion years. For full details of the survey science goals, survey design, target selection, photometry and spectroscopic observations see \cite{Davies18, Davies21}. 
   
The DEVILS regions were chosen to cover areas with extensive exisiting and ongoing imaging to facilitate a broad range of science. In this work we only use the DEVILS D10 field, which covers the Cosmic Evolution Survey region \citep[COSMOS,][]{Scoville07}, extending over 1.5deg$^{2}$ of the UltraVISTA \citep{McCracken12} field and centred at R.A.=150.04, Dec=2.22. This field is prioritised for early science as it is the most spectroscopically complete, has the most extensive multi-wavelength coverage of the DEVILS fields, and has already been processed to derive robust galaxy properties through spectral energy distribution (SED) fitting, see below.

\begin{figure*}
\begin{center}
\includegraphics[scale=0.65]{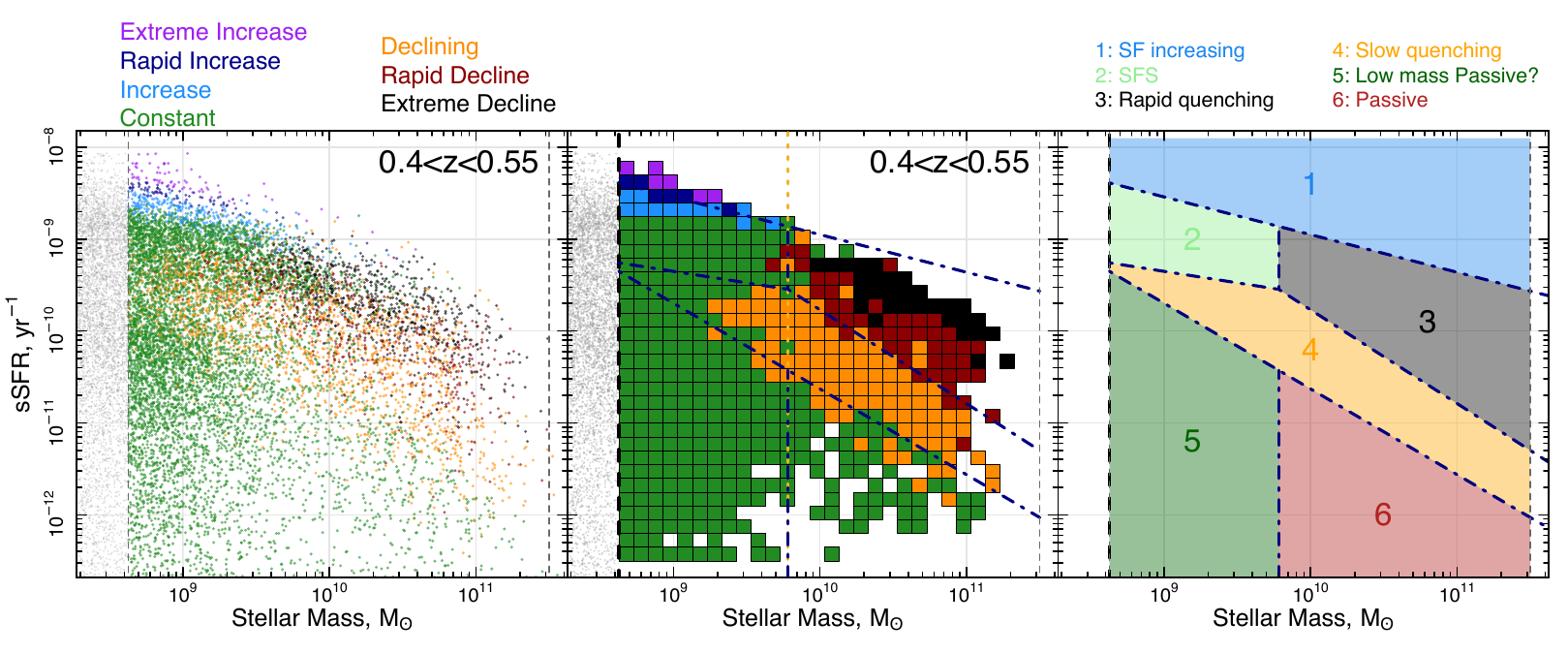}
\caption{Summary of the sample selection undertaken in paper I to identify galaxies with common recent SFHs. Here we only show galaxies at $0.4<z<0.55$. The left panel displays the sSFR-M$_{\star}$ plane, with points colour-coded by the slope of their recent SFH ($\Delta$SFH$_{200\mathrm{\,Myr}}$). These have been split into various bands of $\Delta$SFH$_{200\mathrm{\,Myr}}$ given in the legend above the figure (see paper I for details). Galaxies with common $\Delta$SFH$_{200\mathrm{\,Myr}}$ fall into sub-regions of this plane. To highlight this, the middle panel displays the same data but binned in stellar mass and sSFR. In each bin we show the median $\Delta$SFH$_{200\mathrm{\,Myr}}$ value colour-coded in the same way as the left panel. Using this we then define sub-regions of the plane that bound common recent SFHs, using the navy blue lines. For reference the vertical gold line is the minimum SFR dispersion point at this epoch from D22. In the right panel we label each of these sub-regions based on their recent SFHs. These regions will be used for the discussion outlined in this work. }
\label{fig:summaryP1}
\end{center}
\end{figure*}

\subsection{Sample selection}

In this work we use the outputs of the SED fitting process outlined in \cite{Thorne21} and \cite{Thorne22}. Briefly, \cite{Thorne21} fits galaxies in the D10 region using the \textsc{ProSpect} \citep{Robotham20} SED fitting code to estimate galaxy properties such as stellar mass, SFR, SFH and metallicity. In \cite{Thorne22}, this process is updated to include an AGN model, which allows for the identification of sources hosting bright AGN and improvements to the other derived properties for AGN host galaxies. While the overwhelming majority of sources in the D10 sample do not change their properties, the sources identified as AGN do, in some case, have significant changes, particularly to their SFR and SFH (as UV and MIR-FIR is now attributed to the AGN and not star-formation). A detailed description of how galaxy properties in the D10 sample are affected by the inclusion of the AGN is included in \cite{Thorne22}, so we refer the reader to that work.       

The methodology in both \textsc{ProSpect} analyses uses a skewed log-normal truncated SFH for all galaxies, which assumes that all SFHs are smoothly evolving (the limit of what SFHs from photometric data alone can provide). As such, in this work we will not be able to explore very short duration burst-like SFH variation ($<$100\,Myr). Instead we look for overall global changes to star formation. For example, while we will not capture the true SFH of a galaxy which is currently undergoing a burst of star formation, we will be able to identify that it is increasing in its overall star-formation activity. We also note that for this paper we use the DEVILS-internal D10-\textsc{ProSpect} catalogue \texttt{DR1\_v01}.  Following the \cite{Davies22}, hereafter D22, analysis of the $\sigma_{\mathrm{SFR}}$-M$_{\star}$ relation, we also split this sample into five $\Delta z=0.15$ redshift bins between $0.1<z<0.85$.  These samples are shown in Figure 1 of paper I in this series. 

\subsection{Summary of Paper I and the Validity of ProSpect SFH Fitting}
\label{sec:validity}

\textcolor{black}{In paper I of this series we use the  \textsc{ProSpect}-fitted SFHs from the \cite{Thorne22} work to explore the variation of SFH shapes across the sSFR-M$_{\star}$ plane. We found that there is strong variation in SFH shape as a function of both stellar mass and sSFR. We then defined a simple metric for how galaxies are currently moving through the sSFR-M$_{\star}$ plane using the change in star formation over the last 200\,Myr as:}

\begin{equation*}
\Delta$SFH$_{200\mathrm{\,Myr}}=SFR(t_{LB}=0)-SFR(t_{LB}=200Myr)
\end{equation*}

\textcolor{black}{\noindent where $SFR(t_{LB})$ is taken from the galaxy's \textsc{ProSpect}-derived SFH. This essentially reduces the shape of the recent SFH to a single value. We then show the variation of $\Delta$SFH$_{200\mathrm{\,Myr}}$ across the sSFR-M$_{\star}$ plane and use this variation to define new sub-regions which select for common recent SFHs (Figures 10 and 11 in paper I, and summarised in Figure \ref{fig:summaryP1}). In this paper we will explore the physical properties of galaxies in each of these subregions and use them to determine the mechanisms that are leading to the variation in recent SFHs.}
 
 \vspace{1mm}
 
\textcolor{black}{Before proceeding we must note that one important caveat with this analysis is the validity of \textsc{ProSpect} in robustly fitting the SFH of galaxies in the first place, and by extension $\Delta$SFH$_{200\mathrm{\,Myr}}$. This is covered in some detail in \cite{Robotham20}, \cite{Thorne22} and \cite{Lagos24}, and is also discussed in paper I but we reiterate the key points here. }

\textcolor{black}{In \cite{Thorne22} all galaxies in the DEVILS D10 sample are fit with \textsc{ProSpect} assuming a parametric skewed log-normal truncated SFH and linearly evolving metallically, which is linked to the growth of stellar mass. They use the \cite{Bruzual03} stellar templates, assuming a \cite{Chabrier03} initial mass function. The stellar emission is then dust attenuated and re-emitted in the infrared using the \cite{Charlot00} and \cite{Dale14} models respectively, conserving energy balance. In addition to this stellar/dust component, the galaxy is allowed to contain a flexible AGN component derived from the \cite{Dale14} models with AGN luminosity, viewing angle and optical depth as free parameters.} 

\textcolor{black}{In the literature there is currently a wealth of ongoing discussion with regards to the merits/failings of different SED-fitting approaches in recovering galaxy properties. These arguments largely fall into a comparison of two different fitting methodologies, which use different SFH forms: `parametric' SFHs that fit a smooth parameterised form of the SFH, such as a log-normal or exponentially declining delayed-Tau model \citep[$e.g.$ ProSpect, MagPhys: ][]{Robotham20, daCunha08}, or `non-parametric' SFHs that fit individual time-bins in the SFH, but must enforce some linking between adjacent bins \citep[$e.g.$ Prospector, BAGPIPES:][]{Leja19, Iyer18, Carnall19, Johnson21}.  Opinions regarding the merits of these methods are still divided with significant discussion as the benefits and drawbacks of each approach and specific methodology \citep[$e.g.$][]{Tojeiro07, Dye08, Pacifici12, Pacifici16, Iyer19, Leja19, Carnall19,Lower20, Robotham20, Bellstedt20, Thorne21}. More recently,  \cite{Bellstedt25} performed a detailed comparison of SED fitting using different methodologies, stellar population libraries, stellar atmospheres and metallicity evolutions, and showed that the choice of stellar population library and metallicity prescription, is far more important in robustly deriving galaxy properties than the overall methodology ($i.e.$ the base approach used by different codes, such as parametric vs non-parametric SFHs), likely meaning that both approaches are valid if used with the appropriate models. Here, we will largely ignore this ongoing debate and focus on the validity of \textsc{ProSpect} and the implementation used in \cite{Thorne22} to derive galaxy SFHs.}         
 
\textcolor{black}{Firstly, a number of tests are performed in \cite{Bravo22}, \cite{Bravo23} and \cite{Lagos24} to explore the validity of \textsc{ProSpect} in recovering SFHs from simulated galaxies in the \textsc{shark} semi-analytic model \citep{Lagos18}, when using a very similar methodology to \cite{Thorne22}. They find that, while \textsc{ProSpect} fails to recover some short-timescale burstiness in SFHs, overall the majority of SFHs are well-fit by a skew-log-normal SFH, thus suggesting this is an appropriate choice of functional form for galaxy SFHs.  In paper I we expand upon this to directly compare $true$ $\Delta$SFH$_{200\mathrm{\,Myr}}$ values from simulated \textsc{shark} galaxies, with $\Delta$SFH$_{200\mathrm{\,Myr}}$ values derived from \textsc{ProSpect} fits to the \textsc{shark} photometry. We find that in the majority of cases \textsc{ProSpect} can recover very similar $\Delta$SFH$_{200\mathrm{\,Myr}}$ values between true and fitted $\Delta$SFH$_{200\mathrm{\,Myr}}$ value,} \textcolor{black}{with a median offset between true and \textsc{ProSpect}-fitted $\Delta$SFH$_{200\mathrm{\,Myr}}$ values of 0.01 and mean absolute deviation (MAD) of 0.05.}      

\begin{figure}
\begin{center}
\includegraphics[scale=0.42]{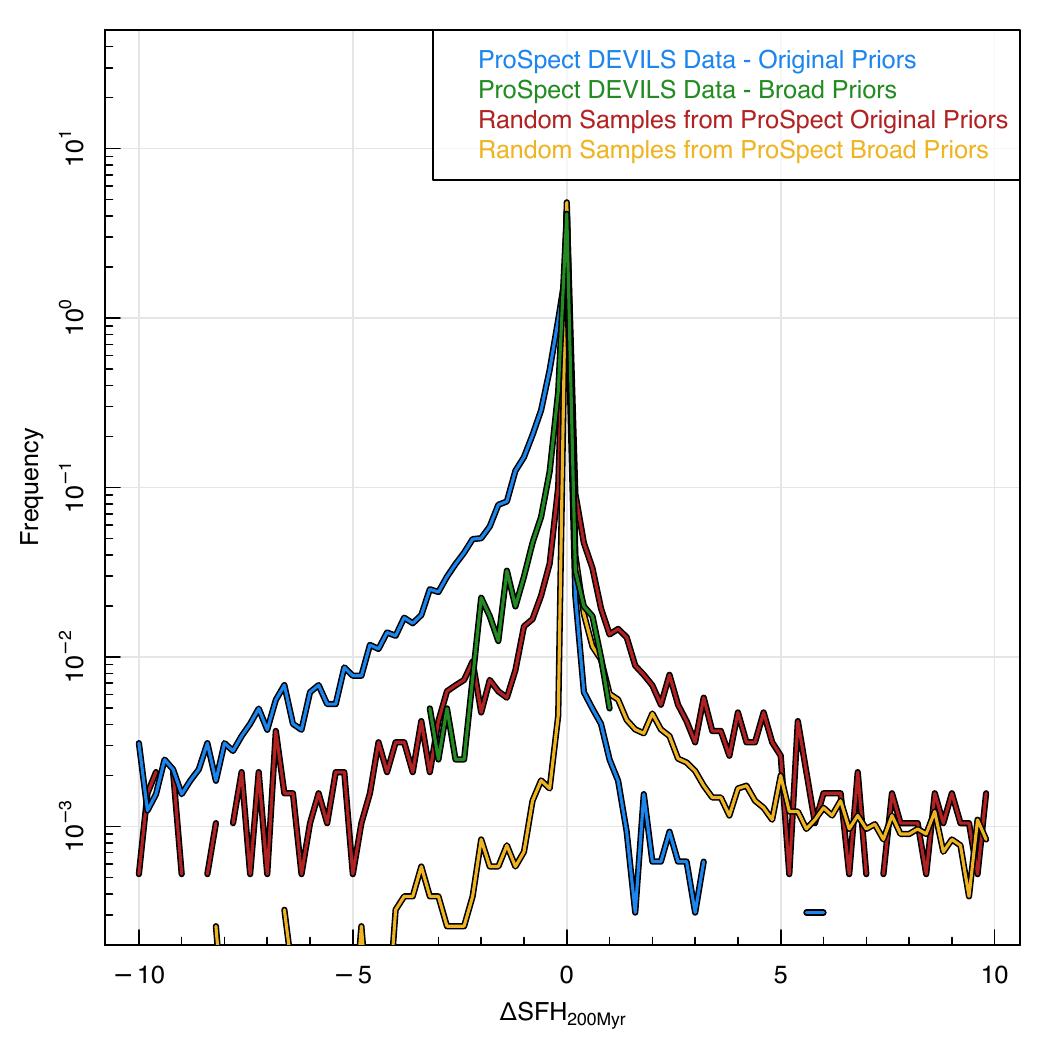}
\caption{Potential impact of choice of \textsc{ProSpect} priors on derived $\Delta$SFH$_{200\mathrm{\,Myr}}$ values. Blue displays the $\Delta$SFH$_{200\mathrm{\,Myr}}$ values for the DEVILS sample, red shows the resultant $\Delta$SFH$_{200\mathrm{\,Myr}}$ values produced from SFHs generated from random sample of the priors described in \citet{Thorne22}, green shows distribution of $\Delta$SFH$_{200\mathrm{\,Myr}}$ values from refits to a random 2000 galaxies in the DEVILS sample using broadened priors\textcolor{black}{, while gold shows $\Delta$SFH$_{200\mathrm{\,Myr}}$ values produced from SFHs generated from random sample with the broad priors (see text for details)}. While the real DEVILS data somewhat traces the distribution from the priors, this is likely not down to any restriction imposed by the priors. The priors in fact go somewhat to allowing a broader range of possible $\Delta$SFH$_{200\mathrm{\,Myr}}$ values to be fit. The peak and distribution of $\Delta$SFH$_{200\mathrm{\,Myr}}$ are largely driven by the global decline of SF in the universe (the majority of galaxies are star-forming and decline with the normalisation of the SFS), and the characteristics of a skew-log-normal SFH (see text for details).}
\label{fig:priors}
\end{center} 
\end{figure}

\textcolor{black}{One other potential source of bias in this analysis is the choice of priors in the \cite{Thorne22} fitting leading to a bias in $\Delta$SFH$_{200\mathrm{\,Myr}}$ values. We note that the priors were selected in \cite{Thorne22} to be appropriate for real galaxy populations observed in the Universe (see the Thorne work for details). However, here we also explore the impact of varying these priors on the final $\Delta$SFH$_{200\mathrm{\,Myr}}$ values obtained. First we take random samples of the SFH parameters within the original \cite{Thorne22} priors, derive SFHs and measure $\Delta$SFH$_{200\mathrm{\,Myr}}$. The distribution of these values are shown in Figure \ref{fig:priors} in comparison to the real values used in this work. From this analysis we find that the prior samples are strongly peaked at $\sim-0.1$ in the same way as the data. They also show a similar spread to low $\Delta$SFH$_{200\mathrm{\,Myr}}$, but extend to much higher $\Delta$SFH$_{200\mathrm{\,Myr}}$ values. This first suggests that the paucity of high $\Delta$SFH$_{200\mathrm{\,Myr}}$ values in the data is not a consequence of the priors - as they are allowable within the prior ranges. However, the peaked distribution may in some way be driven by the choice of priors in the Thorne work. To explore this further, we re-fit a subsample of 2000 sources from the \cite{Thorne22} sample, using} \textcolor{black} {greatly expanded and uniform priors which should not constrain the resultant fits.  These are presented in Table \ref{tab:priors}  in comparison to the original \cite{Thorne22} priors. We also allow galaxy ages and peak SFRs (\texttt{mpeak}) to be greater than the age of the Universe in these new fits, such that we are not constrained by this cosmological assumption. }

\begin{table*}
\caption{Expanded priors used in Section \ref{sec:validity} to test the validity for \textsc{ProSpect} derivation of $\Delta$SFH$_{200\mathrm{\,Myr}}$ values in comparison to original priors used in \citet{Thorne22}.  Here $t_{\mathrm{LB}}$ is the look-back time at the observation epoch of the galaxy.}
\begin{tabular}{ccccccc}
\hline
Parameter & Units & Type & Original Range & Original Prior & Broad Range & Broad Prior \\ 
\hline
\texttt{mSFR} & M$_{\odot}$\,yr$^{-1}$ & Log & [-3,4] & uniform & [-20,20] & uniform\\  
\texttt{mpeak} &  Gyr &Linear & [-2,13.38] & uniform & [-20,20] & uniform \\ 
\texttt{mperiod} & Gyr &  Log & [log$_{10}$(0.3), 2] & 100 erf(\texttt{mperiod} + 2) - 100 & [0,20] & uniform \\ 
\texttt{mskew} & - & Linear & [-3,4] & uniform & [-20,20] & uniform \\ 
\texttt{Zfinal} & - &Log & [-3,4] & uniform & [-20,20] & uniform \\ 
\texttt{maxage} & Gyr & Linear & 13.38-$t_{\mathrm{LB}}$ & fixed & 20-$t_{\mathrm{LB}}$ & fixed \\ 
\hline
\end{tabular}
\label{tab:priors}
\end{table*}%

\textcolor{black} {Based off these broad-prior fits, we then measure the resultant distribution of $\Delta$SFH$_{200\mathrm{\,Myr}}$ values and show this as the green line in Figure \ref{fig:priors}. Finally we repeat the process of taking random samples from these broadened prior distributions, generating SFHs and measuring $\Delta$SFH$_{200\mathrm{\,Myr}}$. The distribution of these samples is shown as the gold line in Figure \ref{fig:priors}. First considering the new prior samples (gold line), we find that the distribution is still peaked at close to zero. This strongly suggests that the peak is not driven by the choice of priors in the fitting, but by the overall methodology using a skew-log-normal distribution. Secondly, we see that many more random samples from these priors are found at larger positive $\Delta$SFH$_{200\mathrm{\,Myr}}$ values. This is expected as the extended priors allow for more sources which are rising in SFRs ($i.e.$ peaking in the future). Considering the new fits using these priors (green line), it is interesting to note that these do not contain the large positive $\Delta$SFH$_{200\mathrm{\,Myr}}$ values allowed by the priors, but are still skewed to negative $\Delta$SFH$_{200\mathrm{\,Myr}}$ values, like the original fits. This suggests that sources with declining SFHs are not driven by the choice of priors. We also find that, in fact, the distribution fit with these new broad priors becomes more peaked, with less extreme values of $\Delta$SFH$_{200\mathrm{\,Myr}}$ than the orginal fits (blue line) - such that the sensible choice of priors actually allows for more extreme variation in $\Delta$SFH$_{200\mathrm{\,Myr}}$. Finally, we also note that in expanding the priors in this way, we significantly reduce the quality of the fits obtained in terms of SED residuals when compared to the DEVILS photometry,  suggesting the choices in \cite{Thorne22} are appropriate.}

 \vspace{1mm}

\textcolor{black}{Following this, it is also useful to note \textit{why} the distribution of $\Delta$SFH$_{200\mathrm{\,Myr}}$ is potentially so strongly peaked at -0.1, if not due to the choice of priors, and why the data has a similar distribution to the priors (particularly on the negative $\Delta$SFH$_{200\mathrm{\,Myr}}$ side). Firstly, the peak position at $\Delta$SFH$_{200\mathrm{\,Myr}}$=-0.1 is likely due to the fact that the majority of galaxies at these epochs are still star-forming and reside on the SFS, and that the normalisation of the SFS (and globally the cosmic SFH) is declining with time. For example, \cite{Thorne21}  find that the normalisation of the SFS declines by $\sim$0.025\,dex every 200\,Myr, so a typical galaxy on the SFS with a SFR$\sim$1-2\,M$_{\odot}$\,yr$^{-1}$ will decline in SFR by $\sim$0.1\,M$_{\odot}$\,yr$^{-1}$ over a 200\,Myr period - exactly the peak in the $\Delta$SFH$_{200\mathrm{\,Myr}}$ distribution. Secondly, the overall distribution of  $\Delta$SFH$_{200\mathrm{\,Myr}}$ values is largely governed by the $\Delta$SFH$_{200\mathrm{\,Myr}}$ values you would obtain from physically-acceptable skew-log-normal distributions - $i.e.$ that the majority of galaxies peak in star formation at some point in the past and have a slow decline at the current epoch. Given that many studies find that log-normal SFHs are well-matched to the galaxy population \citep[in addition to those stated above also, $e.g.$][]{Gladders13, Ciesla17, Diemer17}, we feel this is an appropriate choice.}    
                           
\textcolor{black}{Finally, we also highlight that the use of \textsc{ProSpect} to derive galaxy properties is now covered in many papers \citep[$e.g.$][]{Robotham20,  Bravo22, Bravo23, Bellstedt21, Bellstedt24, Bellstedt25, Thorne21, Thorne22, Thorne23, DSilva23, Cook24}, and is well documented in the literature. In combination, following the above statements, we are argue that there is no evidence to suggest that the methodology used in \cite{Thorne22} is biased and that \textsc{ProSpect} can robustly recover both galaxy SFHs and $\Delta$SFH$_{200\mathrm{\,Myr}}$ values.}

\subsection{Selection of regions with common $\Delta$SFH$_{200\mathrm{\,Myr}}$ values}

While we do not wish to replicate the analysis/figures of paper I here, Figure \ref{fig:summaryP1} provides a summary of the sample selection for each sub-region \textcolor{black}{defined in paper 1} at a single epoch. We note that these subregions evolve with redshift, but in this work we largely combine all epochs (unless otherwise stated). For full details of these selections and how they evolve with redshift, we refer the reader to paper I.  

However, for reference, the sub-regions defined in paper I (and displayed in Figure \ref{fig:summaryP1}) and their characteristics are as follows:\\

\noindent $\bullet$ \textbf{Star formation increasing (Region 1):} Defined to identify sources which have large $\Delta$SFH$_{200\mathrm{\,Myr}}$ values (purple/blue points in the left panel of Figure \ref{fig:summaryP1}) and therefore rapidly increasing SFRs. They are found at high sSFRs (above the SFS) and predominantly at low stellar masses. Their selection boundary does not evolve with redshift, but the fraction of galaxies in this class does evolve very strongly with time, being a dominant part of the population at higher redshifts, but declining to almost no systems as the Universe evolves to $z=0$.  \\

 \noindent $\bullet$ \textbf{Star-forming sequence (Region 2):} Region defined to identify sources which have $\Delta$SFH$_{200\mathrm{\,Myr}}$ values close to zero, but relatively high sSFRs and as such sit on the traditional SFS region (high sSFR green points in the left panel of Figure \ref{fig:summaryP1}). They appear to predominantly fall at lower stellar masses with an upper limit traced by M$^{*}_{\sigma-min}$ \textcolor{black}{(gold vertical line in the middle panel of Figure \ref{fig:summaryP1})}, and form a distinct linear sequence with sSFRs declining slightly with stellar mass. The fraction of galaxies in this class (at the stellar masses probed by this work) also declines with time as the constant SFH and star-forming population moves to lower stellar masses as the Universe evolves (downsizing).  \\
 
 \noindent $\bullet$ \textbf{Rapidly quenching (Region 3):} The upper stellar mass end of the traditional SFS (above M$^{*}_{\sigma-min}$) is found to be populated by galaxies with very negative $\Delta$SFH$_{200\mathrm{\,Myr}}$ values, which are therefore rapidly declining in SFR (black/red points in the left panel of Figure \ref{fig:summaryP1}). This population are likely sources that are undergoing the initial stages of the quenching process and are currently dropping off the traditional SFS. The population does appear to extend to lower stellar masses as the Universe evolves, suggesting that the dominant quenching mechanisms can impact lower mass galaxies over time (see paper I). One of the key focuses of this work will be to explore why this population shows rapidly declining SFRs.       \\

\noindent $\bullet$ \textbf{Slow quenching (Region 4):}  Below the SFS and rapid quenching populations, sits a group of galaxies that all show moderately negative values of $\Delta$SFH$_{200\mathrm{\,Myr}}$, and hence are declining slowly in SFR (orange points in the left panel of Figure \ref{fig:summaryP1}). They occur at all stellar masses probed in this work and their position in the sSFR-M$_{\star}$ plane does not appear to significantly evolve with redshift. As they occur at all stellar masses, they likely represent a quenching pathway that is not dependant on stellar mass, and causes a decline in star formation.         \\
 
\noindent $\bullet$ \textbf{Low mass passive? (Region 5):}   At lower sSFRs than the slow quenching region are galaxies with constant SFHs but very low sSFRs ($i.e.$ passive systems). These are also displayed as green points in the left panel of Figure \ref{fig:summaryP1}. The low sSFR constant SFH population is split into two stellar mass regimes at M$^{*}_{\sigma-min}$. At the low stellar mass end we define the `Low mass passive?' population. It is not obviously clear what type of galaxies this population contains (as discussed in paper I). However, within this work we will explore the properties of these galaxies and attempt to explain their origin.     \\
   
\noindent $\bullet$ \textbf{Passive (Region 6):} Finally, the high stellar mass end of the low sSFR but constant SFH population represents traditional passive systems, and are the end point of the galaxy evolution process.

\section{The physical properties of galaxies with common $\Delta$SFH$_{200\mathrm{\,Myr}}$ }

In this section we use the wealth of information available for the DEVILS sample to explore the physical properties of galaxies in each of our new sub-regions described above, and using bands of $\Delta$SFH$_{200\mathrm{\,Myr}}$ values directly. The aim is to identify any physical property that differentiates the populations of galaxies showing common recent SFHs, potentially linking them to the root cause of the change in the host galaxy.  We initially aim to be \textcolor{black}{completely} agnostic to these properties and explore many available possible correlations with properties \textcolor{black}{such as metallicity, morphology, structure, compactness and the presence of an AGN}. Given our main focus is the processes that cause galaxies to move through the sSFR-M$_{\star}$ plane, we pay particular attention to properties that set apart the star formation increasing and rapidly quenching populations. \textcolor{black}{In summary, we find that the only property that shows a correlation with SFH shape is the presence of an AGN. As such, for ease of reading we relegate the detailed discussion of other physical properties to the appendix and only summarise the key results in the main body of the paper.}

\begin{figure*}
\begin{center}
\includegraphics[scale=0.24]{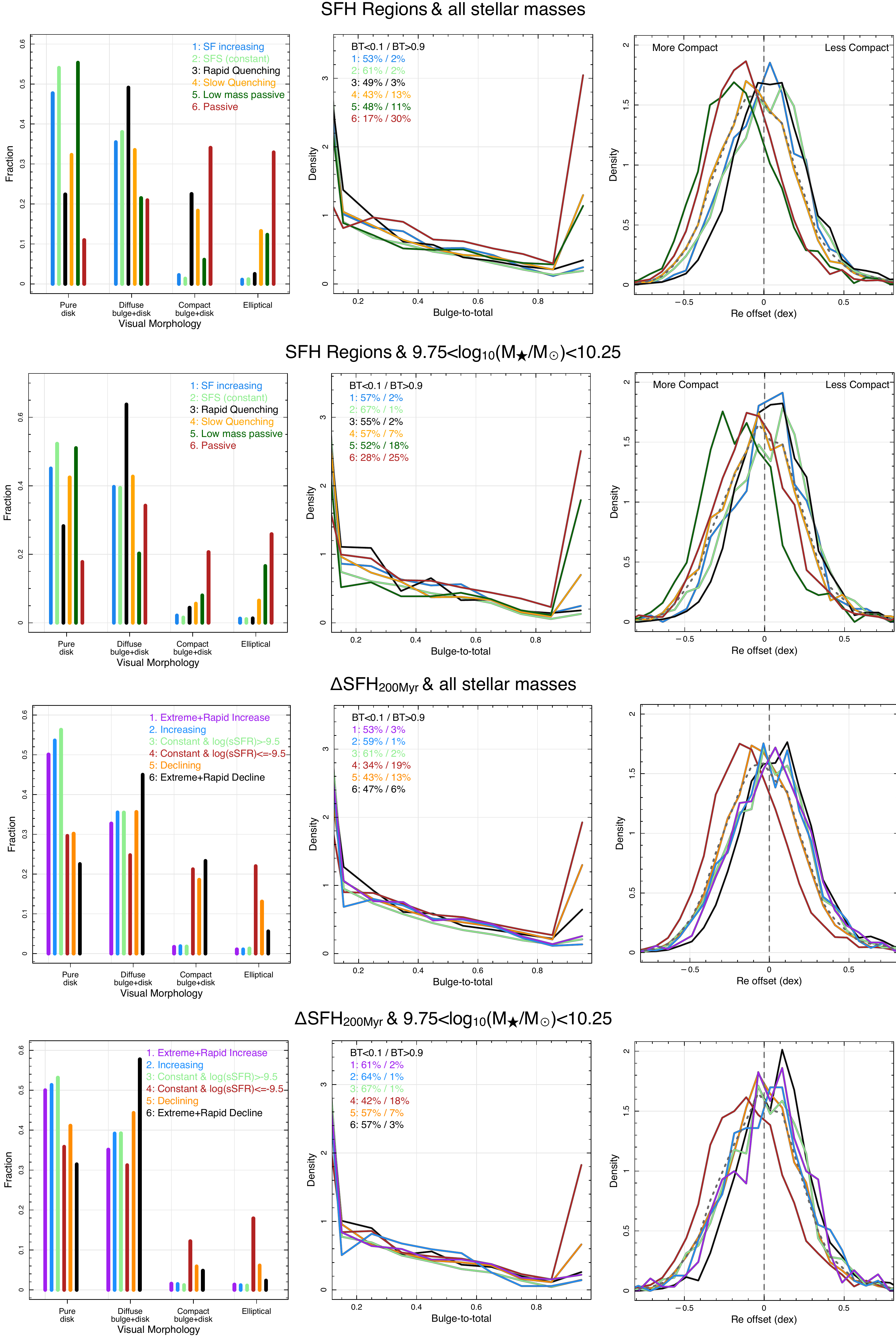}
\vspace{-2.5mm}
\caption{Comparison of the distribution of morphological/structural physical properties of galaxies selected to have common SFHs. Top row: the full sample with common SFHs selected using the region in  the right panel of Figure \ref{fig:summaryP1}, middle-top row: the same but for a narrow mass range at 9.75$<$log$_{10}$(M$_{\star}$/M$_{\odot}$)$<$10.25, bottom-middle: the full sample but with common SFHs selected using measured $\Delta$SFH$_{200\mathrm{\,Myr}}$ values directly, and bottom row: the same but for a narrow mass range at 9.75$<$log$_{10}$(M$_{\star}$/M$_{\odot}$)$<$10.25. Left column shows the visual morphology distribution from \citet{Hashemizadeh21}, middle column the  bulge-to-total distribution from \citet{Cook25},  and right column the R$_{e}$ offset (compactness) distribution (see text for details). Lines in the top two rows are coloured by the common SFH regions outlined in paper I and the right panel of Figure \ref{fig:summaryP1}. Lines in the bottom two rows are coloured by bands of $\Delta$SFR$_{200\,Myr}$ used directly. }
\label{fig:regionComp}
\end{center}
\end{figure*}

\subsection{Properties not strongly associated with common SFHs}

\textcolor{black}{Here we briefly describe the galaxy properties explored in this work that \textit{do not} appear to be strongly correlated with common recent SFHs. For further details we refer the reader to the Appendix. A summary of these results are presented in Figure \ref{fig:regionComp} using: i) the full sample and common SFHs selected using the regions defined above ($i.e$ the right panel of Figure \ref{fig:summaryP1} - top row), ii) a narrow stellar mass range at 9.75$<$log$_{10}$(M$_{\star}$/M$_{\odot}$)$<$10.25 \textcolor{black}{(hereafter the stellar mass-limited sample)}, where all SFH regions overlap and common SFHs selected using the region defined above (middle-top row), iii) the full sample and common SFHs selected using measured $\Delta$SFH$_{200\mathrm{\,Myr}}$ values directly (bottom-middle row), and iv) a narrow stellar mass range at 9.75$<$log$_{10}$(M$_{\star}$/M$_{\odot}$)$<$10.25 and common SFHs selected using measured $\Delta$SFH$_{200\mathrm{\,Myr}}$ values (bottom row). We display both common SFHs selected regions and $\Delta$SFH$_{200\mathrm{\,Myr}}$ values directly, such that the reader can assess that using either methodology does not significantly affect our conclusions. For the exact banding of $\Delta$SFH$_{200\mathrm{\,Myr}}$ values used here, we refer the reader to paper I. }

\textcolor{black}{We also opt to show samples at a narrow stellar mass range as many of the properties we study in this work are correlated with both stellar mass and SFR. To control for this, we would ideally like to use a stellar mass-matched sample. However, given that the regions selected to contain common SFHs do not overlap in stellar mass ($i.e.$ there is no point where regions 2 and 3 or 5 and 6 in Figure \ref{fig:regionComp} overlap), this can not be done. Thus we opt for a small stellar mass range where the regions meet. We also note that in many figures in the appendix, we display the full distribution of these properties in the sSFR-M$_{\star}$ plane, allowing the reader to explore any underlying trends with stellar mass. Where possible in later analyses in this paper, we do control for stellar mass.}

\subsubsection{Metallicity}

First, we simply note that we explored the gas-phase metallicity (Z$_{gas}$) for galaxies in each region, where Z$_{gas}$ is taken from the \textsc{ProSpect} analysis using the \textsc{Z\_final} fitted parameter. The description of Z$_{gas}$ values for DEVILS, their derivation and validity is described extensively in \cite{Thorne22}, and as such we do not discuss this here. \textcolor{black}{However, we note that the \cite{Thorne22} \textsc{Z\_final} measurements for the sample used in this work have typical errors of $<20\%$.} We find no evidence of any correlation between galaxies with common SFHs and their Z$_{gas}$ value (outside of the standard stellar mass-metallicity relation). Thus, we do not find that metallicity is likely correlated with recent change in SFH (at least given the data available to us here), and do not explore this further.

\subsubsection{Morphology}

\textcolor{black}{Next we explore the visual morphological distribution of the galaxies for each of our SFH regions, using the visual classifications of DEVILS galaxies using HST data, presented in \cite{Hashemizadeh21}. A detailed description of this analysis is given in Appendix \ref{sec:Morph}. In summary,}  \textcolor{black}{we do see differences in the distribution of recent SFHs as a function of morphological type (Left column, Figure \ref{fig:regionComp}). Pure disk and diffuse bulge+disk systems are largely dominated by galaxies with increasing or constant SFHs, while elliptical galaxies show predominantly passive SFHs. However, there is a large spread in recent SFH shape in each of the morphological classifications, suggesting that there is no strong correlation between morphology and recent SFH. Moreover, galaxies which have rapidly declining SFHs are found with all morphological types, suggesting that these rapid quenching events are not limited to a specific morphological class. }

\subsubsection{Structure}

\textcolor{black}{Next we explore the bulge-to-total ratio of galaxies in our sample using the structural decomposition of DEVILS galaxies using HST imaging presented in \cite{Cook25}. A detailed description of this analysis is also given in Appendix \ref{sec:Morph}. In summary, we find little difference in the bulge-to-total distribution of galaxies with different SFHs (middle column Figure \ref{fig:regionComp}). In this figure we also show the fraction of galaxies at the extreme ends of the distribution in the legend. Most notably, the `rapidly quenching' (black) population has  a similar bulge-to-total ratio distribution to the `SF increasing' (blue) and `SFS' populations (median $\pm$ standard deviation for B/T of $0.07\pm0.26$, $0.03\pm0.25$, and $0.1\pm0.26$ for SF increasing, SFS and rapidly quenching systems respectively). This suggests that the presence and strength of a bulge is likely not driving these galaxies to vary significantly in SFH. }

\subsubsection{Compactness}

\textcolor{black}{Finally, we also define a metric for the compactness of the galaxy light profile (R$_{e}$ offset, see Appendix \ref{sec:Morph} for details) - which is defined by the size of the galaxy with respect to the typical galaxy at its stellar mass and morphological class. The distribution of R$_{e}$ offset for each SFH type is shown in the right column of Figure \ref{fig:regionComp}. Interestingly, in Figure \ref{fig:ReSel}} \textcolor{black}{we do see a trend in compactness with sSFR, with lower sSFR galaxies having more compact light profiles. However, as discussed in Appendix \ref{sec:Morph}, this is potentially due to disk fading. Considering the right column of Figure \ref{fig:regionComp}, qualitatively we find that rapidly quenching and SF increasing galaxies both show a similar distribution in compactness to SFS galaxies. However, performing a KS test between the R$_{e}$ offset values for rapidly quenching and SFS galaxies we do obtain a p-value=0.0007, suggesting that the distributions are statistically different, with rapidly quenching galaxies typically being slightly more diffuse than their SFS counterparts. However, this effect is very small. The difference in median R$_{e}$ offset values for each sample is $\sim0.02$, while the standard deviation is over 10 times larger ($\sim0.3$). As such, there is considerable overlap in the populations, suggesting that they have no clear separation in R$_{e}$ offset values, indicating that we do not find that rapidly quenching galaxies display any difference in compactness to systems with constant SFHs.  We do note that this is potentially inconsistent with some previous works, which find that recently quenched galaxies at higher redshifts (defined as `post-starburst') are more compact than older, but similar stellar mass galaxies \citep[$e.g.$][]{Whitaker12b,Yano16, Maltby18} - suggesting potentially there was a compaction associated with the quenching event. However, we also indicate that the sample we are focusing on here are not post-starburst galaxies, and would likely not be selected as such. In terms of colour and SFR they appear like SFS galaxies, but only reveal their decline in SF through SED fitting to derive SFHs. Post-starburst galaxies would likely already reside in the populations below the SFS, which do show more compact light profiles.  }

\begin{figure*}
\begin{center}
\includegraphics[scale=0.235]{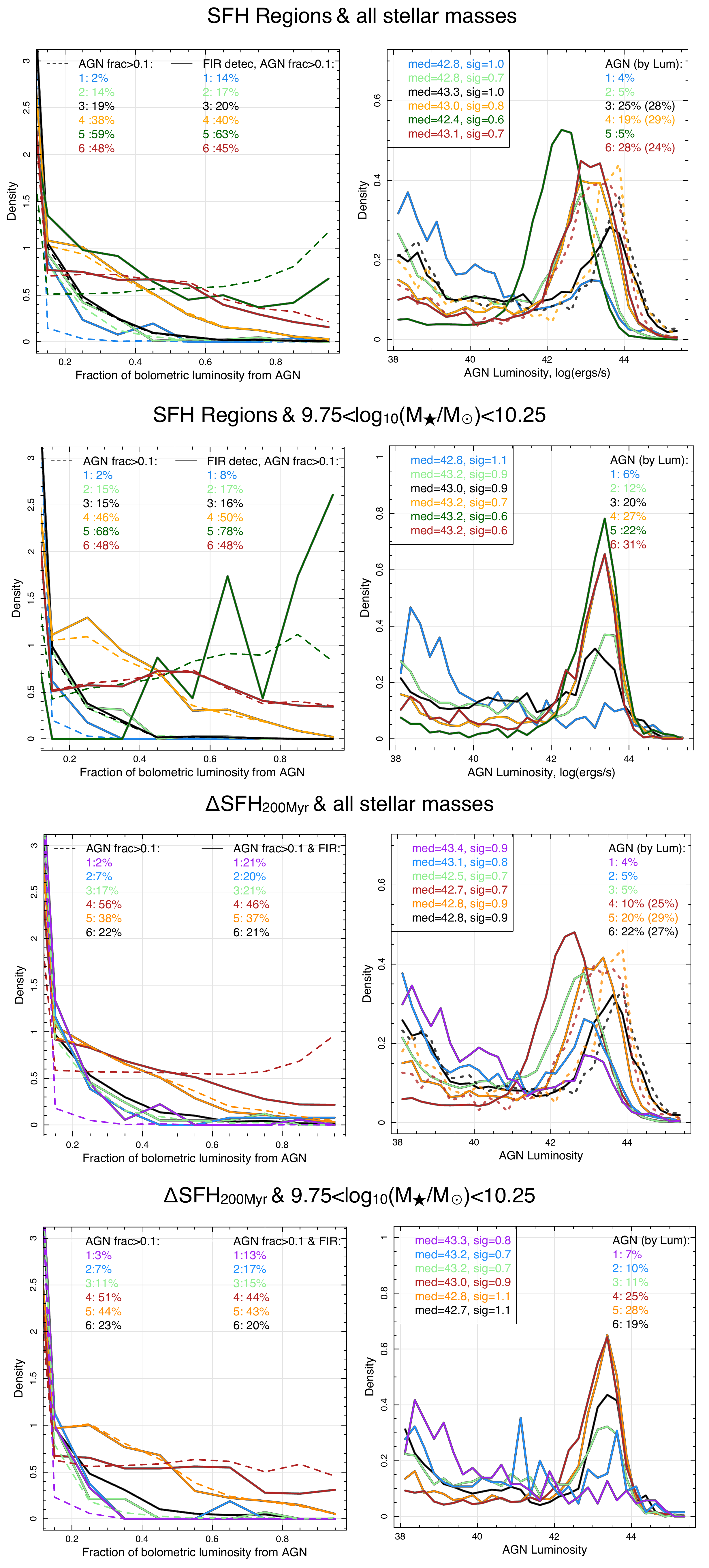}
\vspace{-2.5mm}
\caption{Comparison of the distribution of \textsc{ProSpect}-selected AGN properties of galaxies selected to have common SFHs. Top row: the full sample with common SFHs selected using the region in  the right panel of Figure \ref{fig:summaryP1}, middle-top row: the same but for a narrow mass range at 9.75$<$log$_{10}$(M$_{\star}$/M$_{\odot}$)$<$10.25, bottom-middle: the full sample but with common SFHs selected using measured $\Delta$SFH$_{200\mathrm{\,Myr}}$ values directly, and bottom row: the same by for a narrow mass range at 9.75$<$log$_{10}$(M$_{\star}$/M$_{\odot}$)$<$10.25. Left column shows the AGN bolometric fraction \citet{Thorne22}, right column shown the AGN bolometric luminosity \citet{Thorne22}.  In the left column dashed lines show the full sample, while solid lines show only galaxies with a FIR detection (and therefore well-constrained AGN). In the right column, solid lines show the full sample, while dotted lines show a high stellar mass selection at 10.5$<$log$_{10}$(M$_{\star}$/M$_{\odot}$)$<$11. Fractions of galaxies for various selections are given in the legend. In the top two rows lines are coloured by the common SFH regions outline in paper I and the right panel of Figure \ref{fig:summaryP1}. Lines in the bottom two rows are coloured by bands of $\Delta$SFR$_{200\,Myr}$ used directly. }
\label{fig:regionCompAGN}
\end{center}
\end{figure*}

\vspace{2mm}

In summary, we do see the \textcolor{black}{well-known morphological and structural differences across the sSFR-M$_{\star}$ plane \citep[$e.g.$][]{Kauffmann03, Morselli19, Dimauro22}}, with likely more evolved, higher stellar mass and lower sSFR systems tending from pure disk and diffuse bulge+disk morphologies to compact bulge+disk and elliptical morphologies, and compactness increasing with lower sSFRs. However, given the fact that the most rapidly quenching galaxies show little morphological or structural difference with constant SFH galaxies, it is unlikely that these morphological and structural changes occur prior to the quenching event, and in the case of compactness may just be an observational consequence of disk fading.

\subsection{AGN and their association with common SFHs} 
 
In this next section, we explore the correlation between AGN activity and recent SFH. We select AGN based on a number of different methods and compare to galaxies with common SFHs, both from our selection regions and $\Delta$SFH$_{200\mathrm{\,Myr}}$ values directly. As above, some of the detailed analysis in this area are included for completeness, but relegated to Appendix \ref{sec:AGNapp}. Here we once again only present the key results.

\subsubsection{AGN selected via ProSpect SED fitting}  

\textcolor{black}{First we select potential AGN in our sample using the \textsc{ProSpect} SED fitting outputs calculated in \cite{Thorne22}. This can be done using two approaches. First, identifying sources where a significant \textit{fraction} of their bolometric luminosity arises from the AGN component in their SED fit (called AGN fraction, and described in Appendix \ref{sec:AGNfrac}), and second using the \textsc{ProSpect}-derived AGN bolometric luminosity from the SED fit (described in Appendix \ref{sec:AGNLum}). The former can potentially detect very low luminosity AGN in low stellar mass galaxies, but may be prone to mis-identification of AGN in low SNR sources, while the later will only identify the most luminous AGN, but will likely form a higher purity sample. Here we explore both approaches.}       

\textcolor{black}{In the left column of Figure \ref{fig:regionCompAGN}, we show the distribution of AGN fraction for galaxies in each of our selections for common SFHs as dashed lines. We also note the percentage of galaxies in each region that have AGN fraction $>0.1$ in the top left legend \citep[$i.e.$ galaxies defined as having an  AGN in][]{Thorne22}. The solid lines show the same but only including galaxies where FIR detections are available for their SED fitting (which likely have more robust measurements of AGN fraction). The full analysis of these results is described in Appendix \ref{sec:AGNfrac}.}  \textcolor{black}{We first indicate that the trends are qualitatively the same when considering the full sample (top and third rows), or just stellar mass-limited range at $10^{9.75}<\mathrm{M}_{\star}/\mathrm{M}_{\odot}<10^{10.25}$ (second and bottom rows). For example, the distributions of galaxies with AGN fraction $>0.1$ in each population are visually similar between the top and second row of the left column, with high percentages of populations of passive and slowly declining galaxies having a high AGN fraction in both panels, and the converse true for the SFS and burst populations. This suggests that any trends present here are not solely driven by the different stellar mass distributions in each population.}        

\textcolor{black}{Next, we highlight here that rapidly quenching galaxies (black) show similar AGN fraction distributions to SFS galaxies, where for the FIR-detected sample KS tests between between rapidly quenching and SFS galaxies give a p-value=0.4604 for the full sample, and p-value=0.9792 for the stellar mass-limited sample. This suggests there is likely no strong correlation between low-luminosity AGN, and the astrophysical processes that are driving the quenching/starburst event. We do see large AGN fractions in passive galaxies at  high stellar masses (red lines), which is consistent with previous trends observed for the AGN population \citep[$e.g.$][]{Vitale13, Bongiorno16, Vietri22, DSilva23}, } \textcolor{black}{However, this is also seen for low stellar mass passive galaxies (dark green line). This is somewhat surprising  and is discussed at length in Appendix \ref{sec:AGNfrac}. We do not include this discussion here, as it is not significant to the science question at hand. }  

\textcolor{black}{However, this AGN fraction says little about the overall AGN emission strength (and the likely impact on feedback processes) as it is relative to the overall light distribution in the galaxy ($i.e.$ a very low sSFR galaxy with a weak AGN could have a high AGN fraction value here).} To explore this further, we next consider the distribution of \textsc{ProSpect} AGN luminosities (right column of Figure \ref{fig:regionCompAGN}). In the top-right legend of each panel, we show the fraction of AGN in each sample, where an AGN is defined by an AGN luminosity brighter than the redshift-dependant completeness limits outlined in \cite{DSilva23} - this is discussed in detail in Appendix \ref{sec:AGNfrac}. \textcolor{black}{ First, we note that the distributions here are bimodal for some populations, particularly for the more star-forming galaxies.  Galaxies in the lower AGN luminosity peaks at $<10^{41}$\,ergs/sec, do not have a luminous AGN component in their resultant SED fit (\textsc{ProSpect} simply fits an underlying AGN component which does not strongly contribute to the light distribution) and would not be selected as an AGN based on the limits from \cite{DSilva23}. As such, we ignore them in our subsequent comparisons by limiting to galaxies with AGN Luminosity $>10^{41}$\,ergs/sec.  In the top left legend, we then show the median and standard deviation of each population above $>10^{41}$\,ergs/sec.} 

\textcolor{black}{First, we indicate that we do see some differences when comparing the full sample to the stellar mass-limited range at $10^{9.75}<\mathrm{M}_{\star}/\mathrm{M}_{\odot}<10^{10.25}$, as indicated by median and standard deviations. Here we see that, in the full sample the populations have significantly different median AGN luminosities, with rapidly quenching (black) galaxies showing the highest median AGN luminosity. However, when considering the stellar mass-limited sample, this trend is removed and the populations all have similar median AGN luminosities, suggesting  there is no clear difference in each population's AGN luminosities at a fixed stellar mass in terms of median and standard deviation - potentially indicating that there is not a strong correlation between the luminosity of an SED-selected AGN and recent SFH \textit{at these stellar masses}. However, this is a somewhat difficult comparison as the stellar mass-limited sample, contains very small sample sizes for some populations. In addition, by selecting a stellar mass range where all populations overlap, we are not including the most massive galaxies - where the impact of AGN is likely the strongest. What we do see however, is that the fraction of each population containing an AGN (based on the selection limits of \cite{DSilva23}, given in the top right legend), are similar between the full sample and the stellar mass-limited sample, with SFS and burst populations having low percentages of AGN, and passive and quenching populations having higher percentages of AGN. This somewhat tentatively hints that galaxies which have declining or passive SFHs are more likely to host an AGN, than those bursting or on the SFS - even when selected at lower stellar masses. }

\textcolor{black}{Next, taking just the full samples,} \textcolor{black}{we also see that galaxies in the rapidly quenching region (black lines) have both a high percentage of AGN and host the most luminous AGN. The same is true (but to a lesser degree) in the slow quenching and passive regions, where there are higher fractions of AGN and higher AGN luminosities than for the other regions. However, it is worth highlighting that this could again also simply be a trend with stellar mass, as the selection regions with the highest AGN fractions/luminosities, are also those at the highest stellar masses.} 

\textcolor{black}{To partially explore this, we first also show the same distributions but for only higher mass galaxies at $10^{10.5}<\mathrm{M}_{\star}/\mathrm{M}_{\odot}<10^{11}$ as the dotted lines (fractions in brackets of legend).} \textcolor{black}{While we can not explore this stellar mass range for all of the populations, the rapidly quenching, slow quenching and passive galaxies do overlap here. In this higher stellar mass range, we find that the fraction of AGN goes up in both rapid and slow quenching regions, and the distribution of AGN luminosity are higher (the histograms move to the right).  As such, when only considering higher stellar mass galaxies, the overall result remains the same - that galaxies in the quenching regions of the sSFR-M$_{\star}$ plane have a high fraction of high luminosity \textsc{ProSpect}-selected AGN. This shows more, albeit somewhat circumstantial, evidence that galaxies which are declining in their \textsc{ProSpect} SFH may be associated with an over-abundance of AGN. }

\begin{figure*}
\begin{center}
\includegraphics[scale=0.6]{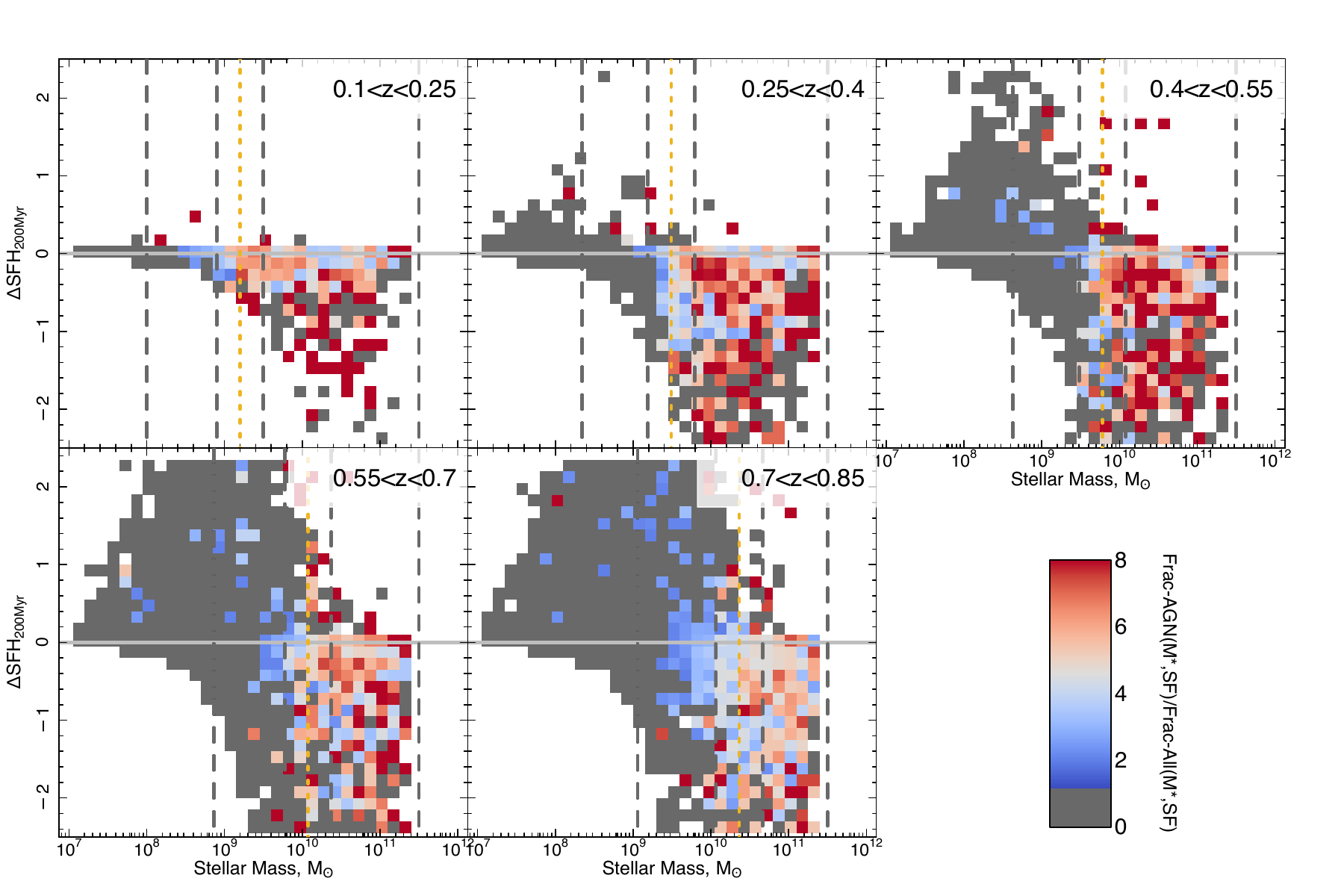}

\caption{Normalised fraction of AGN sources in [M$_{\star}$,$\Delta$SFR$_{200\,Myr}$] bins divided by the normalised fraction of all sources in the  [M$_{\star}$,$\Delta$SFR$_{200\,Myr}$] bins. AGN here are selected to have \textsc{ProSpect} AGN luminosities above the completeness limits derived in \citet{DSilva23}. This figure essentially shows the weighted distribution of AGN in comparison to a random coverage of the full population. $i.e.$ grey bins show where there are few AGN with respect to the total number of galaxies, while colours show where there are many AGN for the total number of galaxies, with the over-density increasing from blue to red. This accentuates where in this parameter space AGN sources differ from the general galaxy population. We find that AGN are found more commonly than expected in the high stellar mass quenching population. Interestingly, we find that this overabundance of AGN is bounded by  M$^{*}_{\sigma-min}$ (gold vertical dashed line) at all epochs.}
\label{fig:AGNcompGyr}
\end{center}
\end{figure*}

\vspace{2mm}

\textcolor{black}{Following the tentative hints suggested above, we then look to explore if AGN are in fact associated with rapidly declining SFHs using a different approach. } We first select \textsc{ProSpect} AGN as galaxies where the predicted AGN luminosity is above the AGN completeness limits outlined in \cite{DSilva23}. We then calculate the number of \textsc{ProSpect} AGN in $\delta$[log$_{10}$(M$_{\star}$)]=0.15 and $\delta[\Delta$SFH$_{200\mathrm{\,Myr}}$]=0.15\,bins, and divide by the total number of AGN in the sample to give the normalised distribution of AGN sources in the $\Delta$SFH$_{200\mathrm{\,Myr}}$ - M$_{\star}$ plane. Next we repeat this process for all galaxies (with or without an AGN) to give a normalised galaxy distribution. Finally we divide the normalised AGN distribution by the normalised galaxy distribution. 

In this approach, bins that have a value $>1$ have an over-abundance of AGN in comparison to the weighted distribution of galaxies across this parameter space. In Figure \ref{fig:AGNcompGyr} we show this distribution, where bins with value $<1$ are shown in grey and those $>1$ are shown in colour with redder colours indicating a larger over-abundance of AGN. Clearly we see that the over-abundances of AGN are found exclusively in the high stellar mass populations that have negative $\Delta$SFH$_{200\mathrm{\,Myr}}$ values (declining SFHs). This is consistent with the results outlined above, but now shown when controlled for stellar mass and the overall distribution of galaxies. This adds more weight to the argument that the AGN are associated with galaxies that show declining SFHs.  

What is potentially more interesting is that the stellar mass range containing an over-abundance of AGN appears to evolve \textit{with} the minimum SFR dispersion point (gold vertical line in each panel). As the minimum SFR dispersion point moves to lower stellar masses as the Universe evolves (as measured in D22), so does the region of over-abundance of AGN, such that the region of over-abundance of AGN always occurs in the population with M$_{\star}$$>$M$^{*}_{\sigma-min}$. This means that the over-abundance of AGN at all redshifts is associated with the high dispersion along the $\sigma_{SFR}-$M$_{\star}$ relation. This is completely consistent with the physical interpretation of the shape of the $\sigma_{SFR}-$M$_{\star}$ relation - that the high SFR dispersion at the high stellar mass end is caused by AGN feedback driving galaxies off the SFS, and the minimum SFR dispersion point occurs where galaxies are too low stellar mass to harbour the massive AGN required to produce significant feedback that would strongly impact a galaxy's SFH \citep[$e.g.$ see][]{Beckmann17, Katsianis19, Davies19a, Davies22, Mulcahey22, Scharr24}. $i.e.$ the minimum SFR dispersion point is actually produced at the point where there are no (or not significant numbers of) strong AGN to drive galaxies away from the SFS and increase SFR dispersion. The results in Figure \ref{fig:AGNcompGyr} are consistent with this picture, once again adding more observational evidence to the argument that AGN feedback drives galaxies off the SFS at high stellar masses.     

\vspace{2mm}

However, care must be taken here as the in-situ sSFR, $\Delta$SFH$_{200\mathrm{\,Myr}}$ values \textit{and} AGN luminosities are all measured from ProSpect. As such, we may be biased towards lower sSFRs by fitting a luminous AGN component. For example, UV and/or MIR flux could erroneously be attributed to an AGN component, increasing the AGN bolometric luminosity, and also decreasing SFRs. \textcolor{black}{Given in Figure \ref{fig:ANGplane} we only find lower sSFR associated with AGN for specific morphologies and at specific stellar masses, this is unlikely to be the case. However,} to try to break some of this degeneracy, in the next section we will explore AGN populations selected completely independently of \textsc{ProSpect}.

\subsubsection{AGN selected via radio and X-ray detections}  

As noted previously, there are potential caveats in using \textsc{ProSpect}-selected AGN to explore correlations between AGN activity and galaxy quenching, when the sSFRs and $\Delta$SFH$_{200\mathrm{\,Myr}}$ values are also \textsc{ProSpect}-derived.  While some of the above analysis suggests this is not the case, we cannot rule out that in fitting an AGN component, \textsc{ProSpect} does not erroneously modify either sSFRs, $\Delta$SFH$_{200\mathrm{\,Myr}}$ or both. For example, light from star-formation at t=0 could be erroneously attributed to an AGN component or vice versa. In addition, even assuming \textsc{ProSpect} is robustly identifying all AGN that produce significant contributions to the UV-FIR broad-band emission from galaxies, this only selects for particular types of AGN at particular stages in their lifetime. This is moderated by AGN duty cycles, observability, and viewing angle \citep[$e.g.$ see discussion in][]{Thorne22}. As such, we now look to other, \textsc{ProSpect}-independent, methods for identifying AGN and compare to the \textsc{ProSpect}-derived sSFR and $\Delta$SFH$_{200\mathrm{\,Myr}}$ values. This will both break some of the degeneracy in purely using our  \textsc{ProSpect} analysis and allow us to build a more complete sample of AGN that covers a broader range of AGN types, duty cycles and viewing angles. 

\vspace{2mm}

First, we use the MeerKAT International GHz Tiered Extragalactic Exploration \citep[MIGHTEE,][]{Jarvis16} radio source catalogues of \cite{Whittam22}, which cover the DEVILS region in COSMOS at 1.4\,GHz. Within the \cite{Whittam22} catalogues they provide radio source identifications as either star-forming galaxy, non-radio-loud AGN, radio loud AGN or unknown (using various selection criteria), and position matched back to optical/NIR sources. We take objects selected as radio loud AGN and carefully position match the optical positions in \cite{Whittam22} to our DEVILS sources to from a sample of 497 radio loud AGN with \textsc{ProSpect} measurements. We do not go into a detailed discussion of the overlap between \textsc{ProSpect}-selected and radio-selected AGN as this is covered extensively in \cite{Thorne22}.

Next, we use the Chandra COSMOS-Legacy survey \citep{Civano16} X-ray point source catalogues of \cite{Marchesi16} to select X-ray AGN in the DEVILS region. This covers sources detected in soft, hard and the full x-ray band of Chandra. \cite{Marchesi16} also provide optical matches to the X-ray point source detections, and as such again simply position match to our DEVILS sources to form a sample of 2148 X-ray bright AGN with \textsc{ProSpect} measurements.

\begin{figure*}
\begin{center}
\includegraphics[scale=0.55]{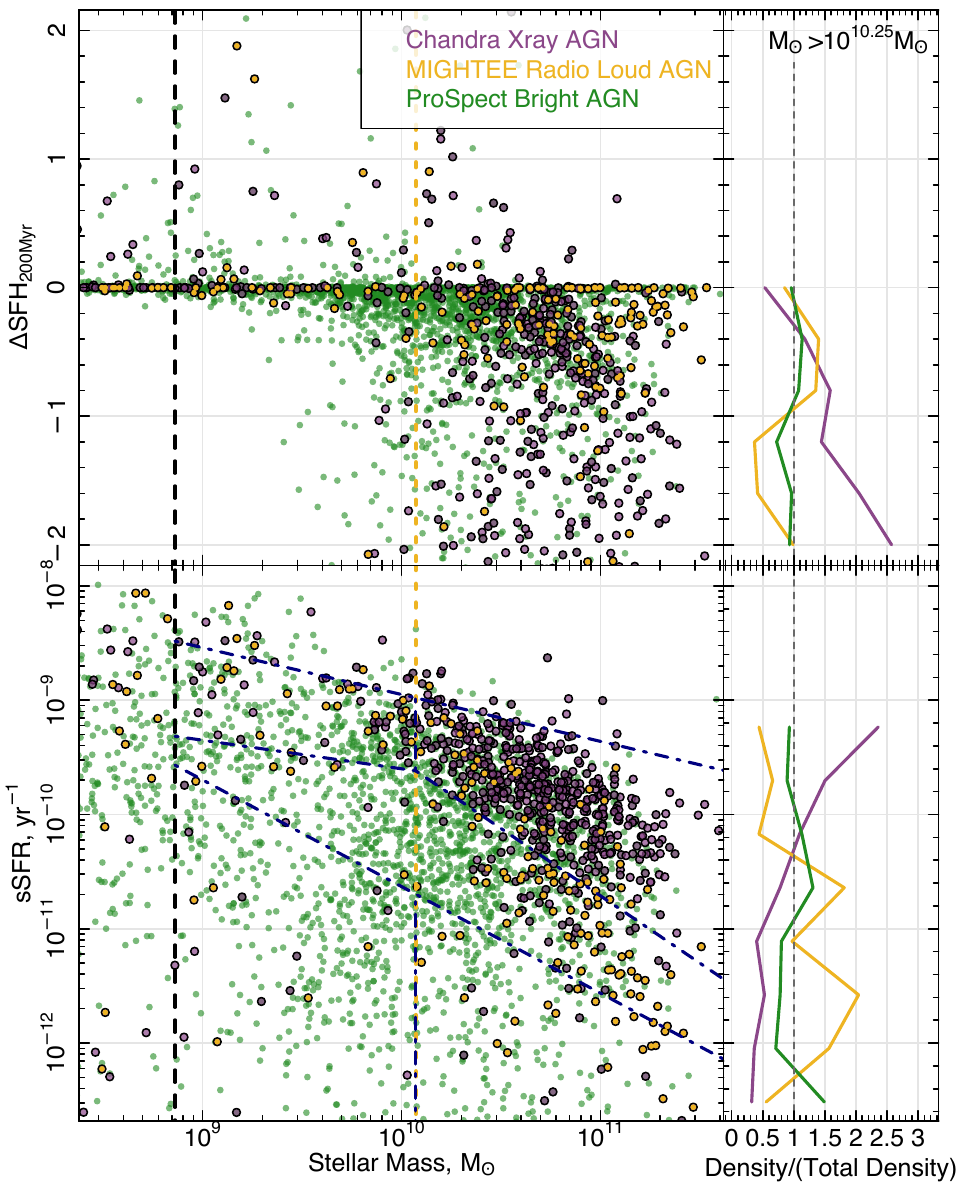}
\includegraphics[scale=0.54]{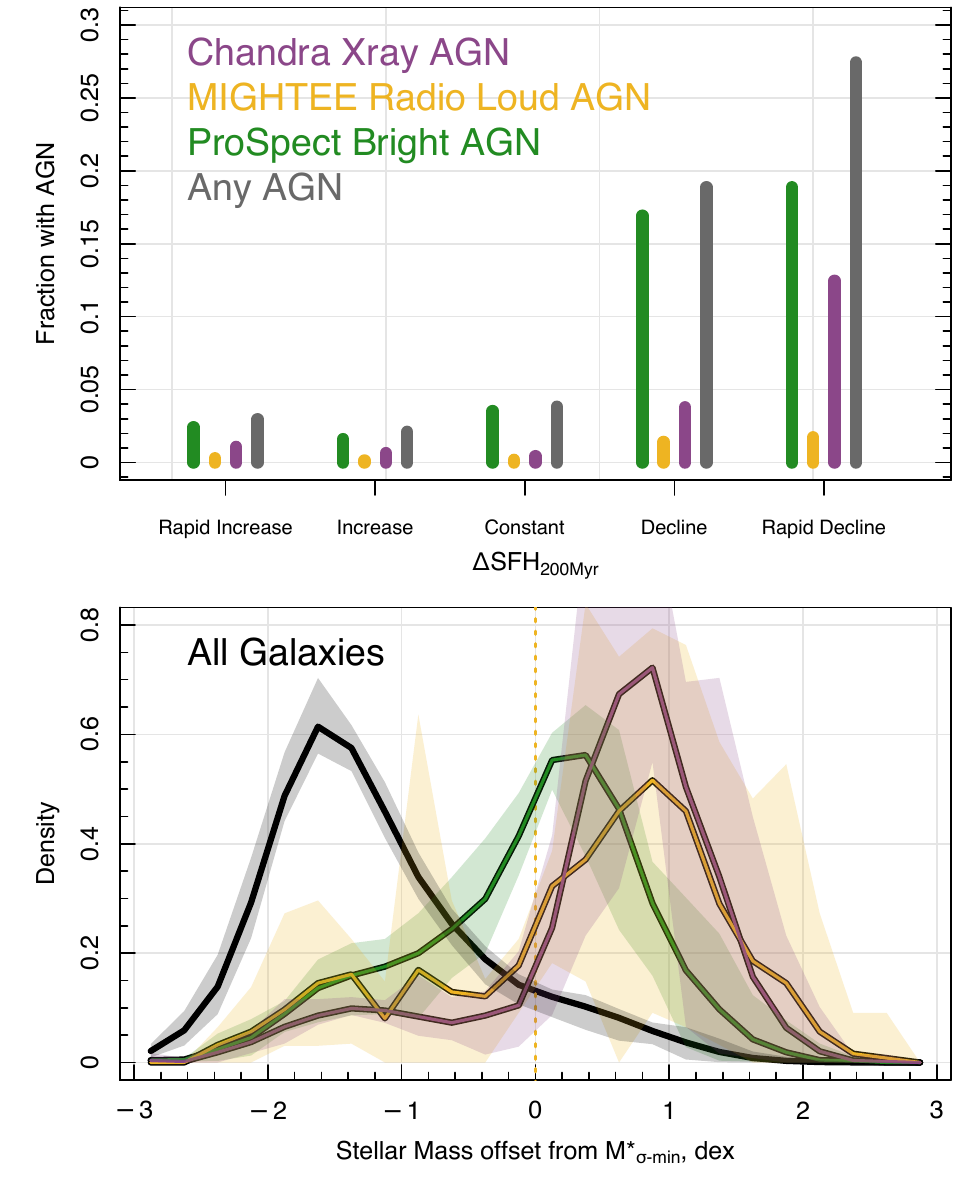}

\caption{Left: The position of AGN selected via various methods at $0.4<z<0.7$ in the $\Delta$SFH$_{200\mathrm{\,Myr}}$ vs stellar mass (top) and sSFR-M$_{\star}$ plane (bottom). We show sources detected as AGN from the \textsc{ProSpect} SED fitting that have an AGN luminosity greater than the limits outlined in \citet{DSilva23} in green, radio loud AGN selected from the MIGHTEE survey \citep{Whittam22} in gold, and X-ray detected AGN from the \textit{Chandra} COSMOS-Legacy survey \citep{Civano16} in purple. Both MIGHTEE and \textit{Chandra} sources are already matched to optical counterparts, see \citet{Whittam22} and  \citet{Marchesi16} respectively, and we then match to our DEVILS sources using a 3\,arcsecond positional match. We highlight the minimum SFR dispersion point with the vertical gold line and in the bottom panel, the selection regions used in this work with navy lines.  The right histograms display the density of the AGN samples as a function of the y-axis, divided by the density of all galaxies, both at M$_{\star}>10^{10.25}$M$_{\odot}$. As such, a value $>1$ displays more AGN than is typical for the general population, and $<1$ less AGN than is typical for the general population. Right top: The fraction of galaxies in each of our $\Delta$SFH$_{200\mathrm{\,Myr}}$ classes that contain AGN from the various selection methods. Clearly galaxies that are declining in their recent SFH show a much higher incidence of AGN. Right bottom: the stellar mass distribution of different AGN samples with respect to the M$^{*}_{\sigma-min}$ point at all epochs. AGN sit above the M$^{*}_{\sigma-min}$ point at all epochs, and in terms of X-ray and radio detected AGN are largely not found below M$^{*}_{\sigma-min}$ at any epoch. }
\label{fig:radioAGN}
\end{center}
\end{figure*}

In the left panels of Figure \ref{fig:radioAGN} we show the position of the \textsc{ProSpect} (green), MIGHTEE radio-loud (gold) and Chandra X-ray-bright (purple) AGN in both $\Delta$SFH$_{200\mathrm{\,Myr}}$-M$_{\star}$ (top) and sSFR-M$_{\star}$ (bottom) planes at $0.4<z<0.7$. In the right histograms of these panels we show the normalised density of AGN, divided by the total normalised density of all galaxies at M$_{\star}>10^{10.25}\mathrm{M}_{\odot}$ (similar methodology to Figure \ref{fig:AGNcompGyr} but in 1D - \textcolor{black}{and therefore at a fixed stellar mass}). In these histograms a value of $>1$ indicates more AGN than would be expected given the general distribution. 

What is immediately obvious from this figure is that different AGN selection methodologies identify sources that occupy different regions of these parameter spaces. \textsc{ProSpect} AGN cover a broad range of stellar masses and sSFRs, but lie predominantly at $>$M$^{*}_{\sigma-min}$ and at slightly lower sSFRs than the SFS (as shown in Figure \ref{fig:ANGplaneAllex}). They also cover a range of $\Delta$SFH$_{200\mathrm{\,Myr}}$ values but with a large number of slowly declining systems at or above M$^{*}_{\sigma-min}$. X-ray-bright AGN fall almost exclusively in our rapidly declining SFH region of the sSFR-M$_{\star}$ plane and also have a very strong over abundance in rapidly declining systems. There also appears to be strong paucity of high stellar mass (M$_{\star}>10^{10.75}\mathrm{M}_{\odot}$) X-ray AGN \textit{without} strongly declining SFHs ($i.e.$ there is a ridge line of X-ray AGN starting at M$_{\star}\sim10^{10.5}\mathrm{M}_{\odot}$ and extending from $\Delta$SFH$_{200\mathrm{\,Myr}}$=0 to more negative values). This potentially indicates that when certain galaxies reach a particular stellar mass, they host an X-ray AGN, and this AGN is almost $always$ associated with declining SFHs. This will be discussed further in subsequent sections. Finally, the radio-loud AGN also predominantly fall in high stellar mass galaxies, but occupy galaxies in the slow quenching or passive region of the sSFR-M$_{\star}$ plane \citep[lower sSFRs than the X-ray AGN. See][]{Moutard20}. This is also consistent with the fact that we find an over-abundance of radio-loud AGN at moderately negative $\Delta$SFH$_{200\mathrm{\,Myr}}$ values, or at constant (but low sSFR) SFHs. 

\vspace{2mm}

While we leave further detailed discussion of these results to the following section, this does suggest that all AGN selections are somewhat associated with galaxies showing a decline in their SFH, just for different populations: \textsc{ProSpect}-selected AGN appear to be linked to both slowly and rapidly declining SFHs at intermediate stellar masses, X-ray-selected AGN with rapidly declining SFHs at high stellar masses, and radio-loud AGN slowly declining SFHs at high stellar masses. To show this in a more straight-forward manner, the top right panel of Figure \ref{fig:radioAGN}, displays the fraction of AGN in our different bands of $\Delta$SFH$_{200\mathrm{\,Myr}}$ (see paper I for full details), where we combine the extreme and rapid bands from paper I into a single `rapid' band. This figure clearly show that irrespective of selection method, AGN are far more prevalent in galaxies with declining SFHs than those with constant or increasing SFHs, such that close to 30\% of rapidly declining galaxies host some form of an AGN. \textcolor{black}{We also note that this result also holds true when only considering the most massive (M$_{\star}>10^{10.5}$M$_{\odot}$) galaxies.}                          

One possible issue with this analysis is that if the \textsc{ProSpect} $\Delta$SFH$_{200\mathrm{\,Myr}}$ measurements are erroneously impacted by the inclusion of an AGN component, $and$ X-ray and radio AGN are more likely to also be detected as an AGN in \textsc{ProSpect}, this may bias our results. $i.e.$ X-ray detected AGN are also fit with an AGN component in \textsc{ProSpect}, and this biases  $\Delta$SFH$_{200\mathrm{\,Myr}}$ measurements. However, as noted previously \cite{Thorne21} also produced the \textsc{Prospect} analyses of our DEVILS galaxies without an AGN component in the fitting process. If we repeat the above process using said fits, we find that for both X-ray-selected and radio-loud AGN there is almost no difference in the positions within each of the planes, and the fraction of galaxies hosting an AGN. As such, the inclusion of an AGN in the \textsc{ProSpect} fitting has no bearing on the coincidence of X-ray-selected and radio-loud AGN and galaxies with declining SFHs. 

Finally, we also wish to explore how the stellar mass distribution of AGN compares to the M$^{*}_{\sigma-min}$ point (as is explored and discussed in Figure \ref{fig:AGNcompGyr}) for different AGN selection methods. To do this using a redshift independent metric, we calculate the stellar mass offset from the M$^{*}_{\sigma-min}$ point for all galaxies at all epochs, and compare the distribution of offsets for different AGN selections. The bottom right panel of Figure \ref{fig:radioAGN} shows this for the full sample (solid lines), and the range of distributions at each redshift independently (shaded regions). We find that while the total galaxy distribution is weighted toward low stellar mass systems, the AGN distribution is weighted to high stellar mass galaxies at all epochs. Interestingly, as in Figure \ref{fig:AGNcompGyr}, we see that galaxies hosting AGN are constrained to stellar masses above M$^{*}_{\sigma-min}$ at all epochs, always coincident with the high SFR dispersion population. This effect is, in fact, even more extreme for X-ray and radio-selected AGN than \textsc{ProSpect} AGN, adding yet more weight to the argument that AGN feedback is the main cause of galaxies moving off the SFS at stellar masses above M$^{*}_{\sigma-min}$ and producing the high SFR dispersion population.

 \begin{figure*}
\begin{center}
\includegraphics[scale=0.65]{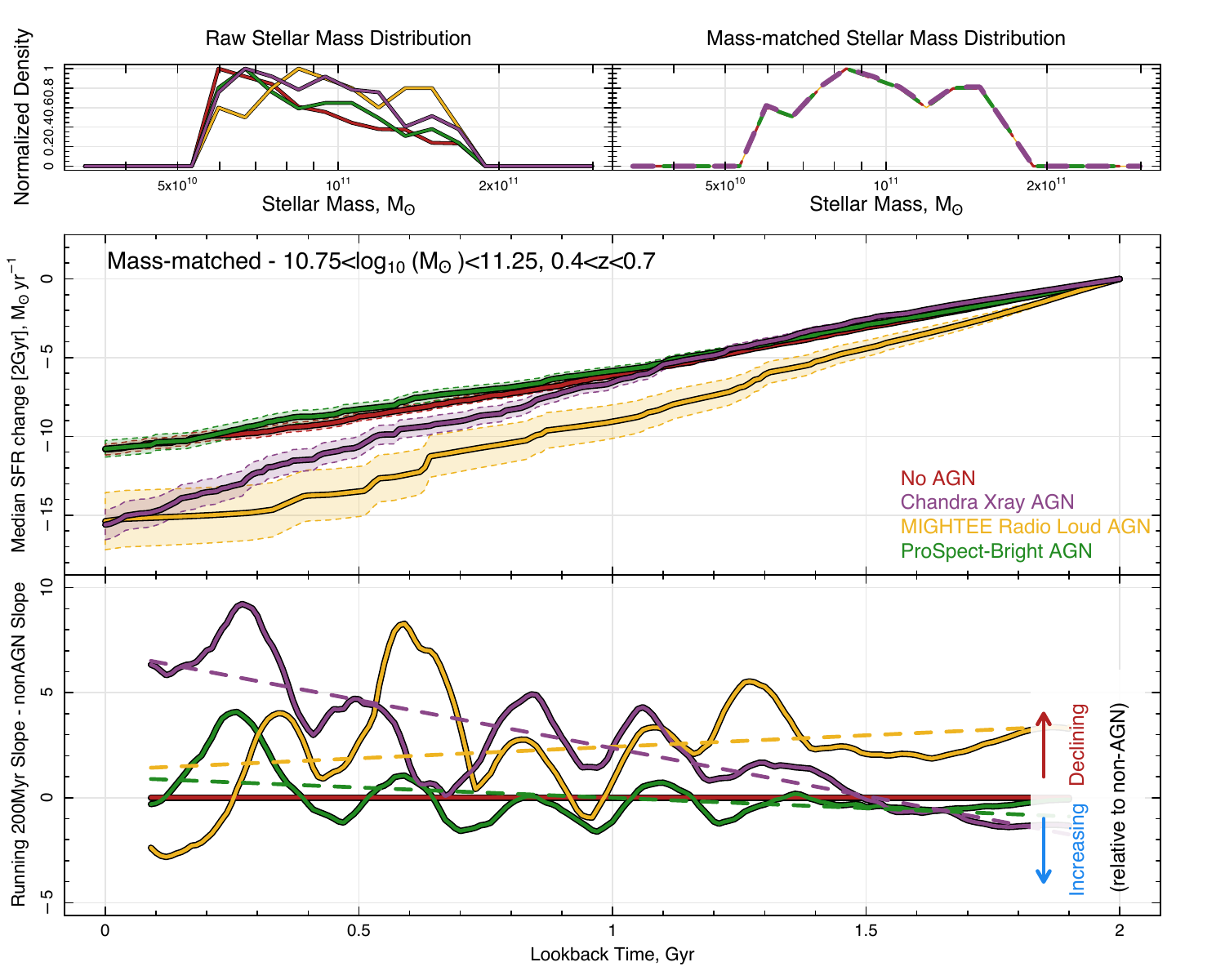}

\caption{The median SFH of galaxies selected at 10.75$<$log$_{10}$(M$_{\star}$/M$_{\odot}$)$<$11.25 and 0.4$<$z$<$0.7, split into non-AGN (red), \textsc{ProSpect}-luminous AGN (green), xray-selected AGN (purple) and radio-loud AGN (gold). The top left panel shows the stellar mass distribution of each subsample. We then derive a stellar-mass-controlled sample for each class, with the resultant distributions shown in the top right. The middle panel show the median recent change in SFH for each subsample over the last 2\,Gyrs (where SFHs are scaled to the same SFR at the look-back time = 2\,Gyr point). The bottom panel shows the running slope of each population in comparison to the non-AGN sample. Dashed lines are a linear fit to these running slopes. }
\label{fig:AGNSFH}
\end{center}
\end{figure*}

\subsection{AGN SFHs in Stellar Mass-Controlled Sample}

\textcolor{black}{As discussed previously, in much of the above analysis it is problematic to produce fully stellar mass-matched samples for comparison. However, now we have identified a potential link between AGN activity and recently declining SFHs we can explore this when controlled for stellar mass. This analysis essentially flips the previous analysis from ``do galaxies with common SFHs show different abundances of AGN'' to ``do galaxies with common AGN show similar SFHs?''. To explore this we first take all galaxies in a narrow stellar mass range 10.75$<$log$_{10}$(M$_{\star}$/M$_{\odot}$)$<$11.25 at 0.4$<$z$<$0.7, where galaxies with all types of AGN classification can be found. We split these samples into galaxies with no evidence of an AGN, a \textsc{ProSpect} luminous AGN,  an x-ray bright AGN or a radio-loud AGN. Note that these sub-samples are non-unique, such that a galaxy can appear in multiple of the AGN classes, $i.e.$ if it is both radio loud and x-ray bright it will appear in both.  The top left panel of Figure \ref{fig:AGNSFH} displays the stellar mass distribution of galaxies hosting AGN using the selections outlined in the previous section. The radio-loud AGN contain the fewest number of galaxies, so we sub-select all other samples to match the stellar mass distribution of the radio-loud AGN. The resultant stellar mass distributions are shown in the top right panel (where all distributions are identical).}

\textcolor{black}{Using this stellar mass-controlled sample, we then take all SFHs in each subsample. We first scale all SFHs to have the same value at a look-back  time of 2\,Gyr. This is a similar process outlined in paper I of this series and is designed to highlight the variation in that recent SFH of galaxies with wildly varying absolute SFRs. Next we take the median SFH for each subsample. The middle panel of Figure \ref{fig:AGNSFH} displays these median SFHs, with error polygons showing the Poisson error in each time-step. We see clear differences between the SFHs of galaxies with different types of AGN. First, we note the non-AGN population (red line), which displays how the typical galaxy at these stellar masses is declining in star-formation. This is the base reference point to compare to the AGN samples. }  

\textcolor{black}{The \textsc{ProSpect} luminous AGN (green line) are very similar to the non-AGN sample, suggesting that these types of AGN are not currently having a strong influence on star-formation in their galaxies. This is consistent with previously presented results, that AGN selected by \textsc{ProSpect} display a broad range of properties and, on average, are relatively similar to the overall galaxy population. Potentially this is due to the fact that SED-fitting can identify lower power AGN than radio or x-ray selections, which have less of an impact on their host galaxies. More interestingly, x-ray selected AGN (purple line) show a SFH significantly different from the non-AGN sample. Over the first Gyr of this 2\,Gyr look-back time, the SFHs of x-ray selected AGN and non-AGN are very similar. However, at $\sim$1\,Gyr in the galaxy's past the SFHs diverge, with x-ray selected AGN rapidly declining in comparison to the non-AGN. This is exactly what would be expected for a galaxy undergoing a quenching event, and is consistent with the previous results presented in this paper. Here, we see clear evidence that x-ray selected AGN are associated with rapidly declining SFHs when controlled for stellar mass. What's more, we also see differences in the radio-loud AGN population (gold). Conversely to the x-ray selected AGN these galaxies deviate from the non-AGN at a much earlier stage, declining in star-formation quite rapidly over the first Gyr and then beginning to plateaux and follow a similar slope to the non-AGN population over the last 0.5\,Gyr.}

 \textcolor{black}{To potentially highlight these differences in slope more clearly, the bottom panel of Figure \ref{fig:AGNSFH} shows the running slope of each of the median SFHs minus the running slope of the non-AGN sample. As such, a positive value indicates that the particular population is declining in SFH more rapidly than the non-AGN sample, while a negative value indicates it is declining more slowly. We also fit the trend of these lines with a linear model (dashed lines) for clarity. This panel highlights the key points from above that: i) the \textsc{ProSpect}-luminous AGN have very similar median SFHs to non-AGN, ii) the x-ray selected AGN initially had similar SFHs to the non-AGN, but in recent times have declined much more rapidly, and iii) the radio-loud AGN were declining more rapidly than non-AGN in the past, but now show similar SFHs.}
 
  \textcolor{black}{These results suggest that x-ray selected AGN may be going through an ongoing quenching event, while radio-loud AGN may have gone through a quenching event at some point in their past, but now show little impact from the AGN on their current SFH. This is discussed further in the following section.}

\section{Discussion}
\label{sec:discuss}

\subsection{Astrophysical mechanisms that drive galaxies off the SFS leading to high sSFR dispersion}

\textcolor{black}{From paper I in this series and our previous analysis here, it is clear that galaxies, which reside in different regions of the sSFR-M$_{\star}$ plane, are evolving very differently in terms of their star-formation and therefore stellar mass growth. This is also seen in numerous other recent works \citep[$e.g.$][]{Sanchez18, Iyer18, Ciesla17, Ciesla21, Broussard19, Emami19, Asada24}.}  In the following discussion we will aim to outline possible physical mechanisms that describe the movement of galaxies through the sSFR-M$_{\star}$ plane leading to the observed $\sigma_{SFR}$-M$_{\star}$ relation and evolution of the SFS (as outlined in Paper I). We start from the premise that the minimum point in the $\sigma_{SFR}$-M$_{\star}$ relation defines the boundary between two different regimes in galaxy evolution processes (as is supported by the results presented in paper I and previously in this paper), and then aim to produce a picture for the astrophysical mechanisms that are occurring in each of these regimes. These mechanisms have been suggested previously by other authors and are described extensively in D22. However, here we aim to provide more direct observational evidence adding weight to these arguments supported by the results in this paper. 

\subsubsection{SFS and self-regulated star-formation at M$^{*}_{\sigma-min}$}

First we argue that underlying the more extreme galaxies that make up the high dispersion populations at either end of the sSFR-M$_{\star}$ plane, there exists a population of galaxies that exhibit constant self-regulated star-formation and fall on a tight sequence that would traditionally be deemed the SFS \citep[$e.g.$][]{Hopkins11, Steinhardt20, Popesso23}. These galaxies typically are found at around sSFR$\sim10^{-9}$\,yr$^{-1}$, and have a shallow decline in sSFR with stellar mass. We show that this population is largely found at or below the M$^{*}_{\sigma-min}$ point, as above this galaxies begin to show rapidly declining SFHs that aren't consistent with a self-regulated mode of star-formation. In fact, it appears that M$^{*}_{\sigma-min}$ occurs at the point were we can truly measure the dispersion of this constant self-regulated star-formation population, without the more extreme populations of galaxies driving up the measured dispersion. Interestingly, the SFS population with constant SFHs does not seem to evolve in sSFR much over the epochs probed in this work. More it is the relative numbers of galaxies in this population that changes, leading to overall changes in the sSFR-M$_{\star}$ plane (as measured in Paper I). This population is predominately composed of pure disk or low bulge-to-total disk+diffuse bulge systems \textcolor{black}{(Figure \ref{fig:BTplane})}, with low AGN fractions  \textcolor{black}{(Figure \ref{fig:ANGplane})}. This is what would be expected for systems undergoing self-regulated star formation, that are not being impacted by either external or large-scale internal processes \citep[$e.g.$][]{Hopkins11}. 

At the M$^{*}_{\sigma-min}$ point we do see a large fraction of galaxies that are declining in SFR which sit below the SFS. They are declining slowly as they transition across what would traditionally be deemed the `green valley' \citep[$e.g.$][]{Wyder07, Schawinski14}. There may be a number of different mechanisms that lead to galaxies declining in star-formation at these stellar masses. First, we do not see strong evidence for morphological differences between the SFS galaxies and declining galaxies here - suggesting the morphology and structure likely do not have a strong impact on driving these galaxies to declining SFR \citep[consistent with][]{Cortese19}. We do find a relatively large incidence of \textsc{ProSpect}-selected AGN at this point \textcolor{black}{(Figure \ref{fig:ANGplaneAllex})}, which appear associated with a reduction in sSFR. As such, at least some of this slow decline in star-formation may be driven by lower luminosity AGN, which are not observable in the X-ray or radio \citep[$e.g.$ see][]{Almeida23}. 

However, one important potential quenching mechanism at these stellar masses, which we do not explore in this work, is the impact of environment. A large fraction of these galaxies will likely be satellites in group scale haloes, which are known to have their star-formation significantly impacted by their environment \citep[$e.g.$][]{Peng10, Wetzel13, Davies19b, Siudek22}. However, we do not currently have environmental metrics for the DEVILS sample and cannot explore this further. These metrics are currently in preparation, and exploring their correlation with quenching mechanisms will be the subject of future work.

\subsubsection{Starbursts and SFH variation below M$^{*}_{\sigma-min}$}
\label{sec:Starbursts}

As we move to lower stellar masses the sSFR dispersion increases significantly.  In previous work \citep[$e.g.$][]{Katsianis19,Davies19a, Davies22} this has been attributed to stochastic star-formation processes caused by star-formation feedback driving galaxies away from the tight self-regulated SFS and symmetrically `puffing-up' the sSFR distribution of galaxies \citep[$i.e.$ see][]{Genel18, Caplar19, Hahn19}. These stochastic processes would be seen as galaxies that are temporarily increasing in star-formation to move above the SFS and temporarily declining in star formation to move below the star-forming sequence, before eventually returning to a self-regulated mode. We must note again here that our \textsc{ProSpect} analysis is not sensitive to very short timescale variations in star-formation. However, it does allow us to identify galaxies that have both increasing or declining star-formation in their recent history ($i.e.$ see paper I). This is evidenced by the galaxies that do show a decline in star-formation at higher stellar masses - and comparisons to the \textsc{shark} semi-analytic model \citep{Lagos18b, Lagos24} where for low stellar mass galaxies, the true SFH for burst/quenching galaxies is robustly recovered (once again, see paper I). 

\textcolor{black}{Assuming that \textsc{ProSpect} can correctly recover galaxies which are either declining or rising in their star-formation in their recent history, the pertinent question is - are our current results from paper I and II of this series consistent with this picture of the low stellar mass end of the SFS symmetrically `puffing-up' through stochastic star-formation events? The answer appears to be - not entirely. While we do see a large number of systems that have rapidly increasing SFHs above the SFS at the low stellar mass end, which is driving much of the dispersion, we find very few galaxies that show declining SFHs below the SFS, What's more we do not see any galaxies below the SFS with increasing SFRs; which we would expect if they are returning to their self-regulated state.  This is clear in $e.g.$ the left panel of  Figure \ref{fig:summaryP1}, where there are no blue/purple points below the SFS, and the majority of the low sSFR galaxies at low stellar masses, have relatively constant SFHs (green points). So it is appears from these results that large sSFR dispersion seen at low stellar mass is potentially a combination of star-bursting-like galaxies above the SFS \citep[blue/purple points, $e.g.$][]{Broussard19, Emami19},  $and$ a much greater diversity in the \textit{normalisation} of the constant SFH sources (large sSFR spread green points).}

\textcolor{black}{This is a potentially very interesting result, which is somewhat in tension with the previous theoretical interpretation of the increased sSFR dispersion at the low stellar mass end being completely due to highly stochastic sources \citep[$e.g.$][]{Katsianis19,Houston23}. Clearly this topic requires further investigation and here we attempt to partially elucidate this picture via two approaches.}

\textcolor{black}{First, we consider if the low stellar mass sources with constant SFH in our sample, are correctly fit by \textsc{ProSpect} and truely have long duration constant SFRs. Full details of this analysis are given in Appendix \ref{sec:starburst}, but to summarise, we find that, within the limitations of our analysis, these galaxies do appear to have constant SFHs across their entire lifetimes and we find no evidence that they have been erroneously fit. Next we once again use the \textsc{shark} semi-analytic model to explore the prevalence of stochastic star-formation vs long-duration constant star-formation at low stellar masses. Details of this analysis are presented in Appendix \ref{sec:shark}, where we define a metric for stochasticity in recent SFHs and find that there is only a very weak trend with stellar mass. We also find a broad spread of stochasticities at all stellar masses - potentially suggesting that galaxies with long duration constant SFHs, can populate the low stellar mass regions of the  sSFR-M$_{\star}$ plane. However, we also find that \textsc{shark} does contain galaxies that sit below the SFS which are rapidly increasing in star-formation (consistent with the previous theoretical interpretations). }

 \textcolor{black}{Obviously, this picture is far from clear and warrants better study to further understand the processes which drive the large sSFR scatter at low stellar masses. However, there is little more that can be stated here due to limitations in our current data/analysis.}

\subsubsection{AGN feedback and quenching above M$^{*}_{\sigma-min}$}

At the high stellar mass end, the sSFR dispersion also increases significantly but the root causes of this are likely very different. At these stellar masses, we find a large variation in SFH shapes, with a strong dominance of declining SFRs pushing galaxies below the SFS and likely leading to the large sSFR dispersion. We find that the rapidly declining population only becomes apparent at M$_{\star}>$M$^{*}_{\sigma-min}$, suggesting that there is a stellar mass dependence on the cause of the rapid quenching \citep[stellar mass quenching?][]{Peng10, Darvish18}, but that this stellar mass dependance may evolve with time. We also see a trend where galaxies that lie on the traditional SFS region at these stellar masses have rapidly declining SFRs, at lower sSFR galaxies have more slowly declining SFRs and lower still there are passive galaxies with $\sim$constant SFHs. Could this be an evolutionary sequence, where galaxies begin quenching rapidly at a particular stellar mass and then slow in their quenching rate until they become passive? Or could this be two different quenching paths (one fast, one slow) caused by completely separate physical mechanisms \citep[$e.g.$][]{Schawinski14, Walters22, Freitas22}? The slow quenching galaxies do appear to be ubiquitous over a broad range of stellar masses, while the rapid quenching galaxies occur only at this high stellar mass end \citep[However, c.f.][]{Walters22}.  This potentially indicates that stellar mass (or a property correlated with stellar mass) is leading to the rapid quenching but that slower quenching mechanisms are driven by a different physical process. Once again, we note that this is largely consistent with the variation in quenching timescales as a function of stellar mass found by $e.g.$ \cite{Bravo23} and  \cite{Schawinski14} and is consistent with the EAGLE simulation results of \cite{Matthee19}. The key question here is: what is causing galaxies to start rapidly declining in star formation at a given stellar mass? 

Many previous studies \citep[][]{Croton06,Fabian12, Bluck22, Scharr24} have suggested that it is likely AGN feedback that suppresses star-formation at the high stellar mass end by heating or expelling gas from the system. In fact, such AGN feedback in now ubiquitous in galaxy evolution simulations, required to produce the overall stellar mass distribution of galaxies in the local Universe \citep{Bower06, Sijacki07, Vogelsberger14, Schaye15, Lagos18}. This feedback would cause galaxies to drop off the SFS resulting in a high sSFR dispersion. However, there are many other proposed smoking guns that could lead to quenching at high stellar masses, such as morphological and structural changes \citep[$e.g.$][]{Faisst17, Wang18,Zolotov15}, or stellar mass in itself initiating a decline in star formation as a galaxy can not accrete enough gas to maintain its high SFR \citep[cosmological starvation][]{Feldmann15, Man18}. Here we will discuss the possible astrophysical mechanisms which could cause these galaxies to rapidly quench and aim to identify the most likely culprit.

\vspace{2mm}

In Appendix \ref{sec:Morph} we show that the rapidly quenching population looks very similar to the SFS population in structural characteristics. They have similar visual morphological classifications (albeit with a larger incidence of diffuse bulge + disk systems over pure disks), similar bulge-to-total distributions and similar compactness. This suggests that the transition from SFS constant SFHs to SFS rapidly declining SFHs is not strongly preceded by a change in morphology or structure, and that morphology/structure is unlikely to be the root cause of galaxy quenching at these stellar masses. This is consistent with previous works \citep[$e.g.$][]{Cortese19, Liu19, Bluck22}, however see \cite{Martig09} for a different perspective. We do however, see morphological/structural differences as galaxies drop further below the SFS and decline in sSFR, with the fraction of compact bulge+disk and elliptical morphologies and bulge-to-total distributions both increasing. As this change is likely not associated with the quenching, the immediate question is why are a lower sSFR and a morphological change coincident? One potential explanation is simply galaxy age. The galaxies with lower sSFRs (particularly at the high stellar mass end) are likely older than galaxies with high sSFR \citep{Ciesla17, Lee18}. Older galaxies have had more mergers and are more likely found in over-dense environments, both strong drivers of morphological and structural transformations. We do also find that galaxies at lower sSFRs have more compact single S\'{e}rsic light profiles at a given morphology and stellar mass \textcolor{black}{(Figure \ref{fig:ReSel})}. As discussed previously, this is likely an observational effect caused by disk fading as a galaxy declines in star-formation (the central region fades less rapidly leading to more concentrated profiles), not an astrophysical change which is somehow driving the decline in star-formation \citep[$e.g.$ see][]{Croom21}.  

In summary we find no observational evidence in our sample that morphological and/or structural changes precede the decline in star-formation, but likely occur after the decline in star-formation has already occurred. Thus, we suggest that these changes are not the driver of quenching at the high stellar mass end.                    

\vspace{2mm}
 
However, the same cannot be said for AGN. In our analysis in this paper, we see a strong correlation between declining star-formation rates and AGN activity in galaxies above M$^{*}_{\sigma-min}$, for multiple different AGN selection methodologies. First we see an over-abundance of luminous \textsc{ProSpect}-selected AGN in disk and diffuse bulge+disk systems at or just above M$^{*}_{\sigma-min}$. These galaxies fall in the slow quenching region of the sSFR-M$_{\star}$ plane and are associated with lower median sSFR than non-AGN systems at the same stellar mass \textcolor{black}{(Figure \ref{fig:ANGplaneAllex})}. This suggests that SED-selected AGN may be driving a moderate decline in star-formation at these stellar masses. Next we see a strong correlation between rapidly declining SFHs and X-ray-selected AGN above M$^{*}_{\sigma-min}$, where a significant fraction of galaxies with rapidly declining SFHs are hosting an X-ray luminous AGN \citep[consistent with the finding of][ for the SIMBA simulations]{Scharr24}. Finally, we find that massive galaxies in the slow quenching region have an over abundance of radio-selected AGN, which fall at lower sSFRs than X-ray AGN at the same stellar masses. We also find that AGN, and particularly those that are associated with a decline in star-formation, are largely restricted to stellar masses above M$^{*}_{\sigma-min}$ at all epochs, even as M$^{*}_{\sigma-min}$ evolves. This means that AGN are both associated with galaxies with declining SFHs \textit{and} always fall in the measured high SFR dispersion region of the sSFR-M$_{\star}$ plane as it evolves. 

In combination these results provide strong observational support for the argument that at the high stellar mass end it is AGN feedback which leads to galaxy quenching \citep[$e.g.$][]{Bluck22, Scharr24, Harrison24}, drives galaxies off the SFS leading to the break point in the SFS \citep[see][]{McPartland19} and increases the SFR dispersion, consistent with previous simulation predictions of $e.g.$ \cite{Katsianis19, Scharr24}. 

What is more, we also see that AGN selected using different observational techniques are associated with galaxies in different regions of the sSFR-M$_{\star}$ plane \citep[as][]{Delvecchio15, Comerford20} and of different strengths of SFH decline (Figure \ref{fig:AGNSFH}). AGN selected using these different techniques likely probe different phases of AGN evolution and/or energy output/mode \citep[see discussion in][]{Whittam22}. AGN selected using X-ray emission but with no associated radio emission are likely in a higher power quasar mode, where large amounts of radiative energy is being added into the surrounding gas, potentially driving it out of the galaxy and suppressing star-formation \citep[][]{Silk98, Pontzen17}. Radio-selected AGN, are likely in a lower radiative power mode but will deposit kinetic energy into the surrounding halo via large radio lobes, effectively stopping the halo gas from cooling and accreting onto the galaxy - suppressing future star-formation \citep[$e.g.$ see reviews of][]{Alexander12, Harrison17}. Conversely SED-selected AGN likely cover a range of different AGN types and emission modes, and may potentially be identifiable over longer periods of the AGN duty cycle as they can identify both obscured and unobscured AGN in various phases \citep[see discussion in][]{Thorne22}. In addition, the deeper optical-NIR data required for SED fitting potentially allows for the identification of AGN arising from much lower mass black holes than the other selection methods. 

These physical interpretations of AGN selected using different techniques are consistent with the results we see here, \textcolor{black}{particularly in the stellar mass controlled sample of Figure \ref{fig:AGNSFH}}. The galaxies that have rapidly declining SFHs are most strongly associated with X-ray selected AGN \citep[$e.g.$][]{Harrison24}. AGN emission in this highly radiative mode will efficiently expel gas reservoirs rapidly suppressing star-formation on short timescales. Similarly, radio-AGN are more strongly associated with slowly quenching and already passive systems, where the radio lobes suppress the accretion of new cool gas, with which to form new stars \citep[$i.e.$ maintenance mode, see][]{Choi18}. This also has the effect of suppressing star-formation, but does not initially expel the gas already residing in the galaxy, leading to a slower quenching process as this gas is consumed. However, care must be taken here as there is a strong correlation between the presence of a radio-loud AGN and stellar mass, such that we may just be finding that the majority of high stellar mass galaxies are declining in star formation, and they also host a radio-loud AGN. \textcolor{black}{However, this does not seem to be the case as in Figure \ref{fig:AGNSFH} we show a stellar mass controlled sample of galaxies and find that radio-loud AGN do show distinctly different median SFHs to other samples.} The SED-selected AGN cover a much broader range of AGN properties, and thus would show both rapid and slowly declining SFHs (as seen here). As such, not only do we find that AGN activity is associated with declining SFHs, but that the rate of this decline is strongest for galaxies with X-ray-selected AGN - consistent with physical models for AGN feedback, where the radiative mode produces strong and rapid expulsion of gas leading to quenching \citep[][]{Silk98, Pontzen17}.                                      

\vspace{2mm}

The final piece of the puzzle in explaining the movement of galaxies through the sSFR-M$_{\star}$ plane, as caused by AGN feedback, is determining what drives the AGN to turn on in the first place - and initiate the quenching process. While we cannot provide any direct evidence here, we do have some tentative observational trends that point towards possible explanations. These will then be explored in later work. Firstly, we observe that galaxies with rapidly declining SFHs and AGN, sit above the M$^{*}_{\sigma-min}$ point at all epochs (Figure \ref{fig:radioAGN}). As M$^{*}_{\sigma-min}$ evolves to lower stellar masses with time, this suggest that the process that triggers an AGN and leads to quenching also moves to lower stellar masses as the Universe evolves \citep[similar to that found in][]{Delvecchio17}. Secondly, we find that galaxies which have X-ray bright AGN and declining SFHs, all tend to occur at $\sim10^{10.5}$M$_{\odot}$. Interestingly, this is also the stellar mass at which the peak in $z=0$ major merger rates occurs \citep[$e.g.$][]{Robotham14}. Many previous authors have suggested that AGN can be triggered by major mergers \citep[$e.g.$][]{Wassenhove12, Blecha18, Ellison19}. As such, it is an intriguing prospect that our declining SFH galaxies with bright X-ray AGN may be associated with post-major-merger systems. However, we cannot yet say anything further about this possibility as we are still in the process of identifying merger systems in DEVILS \citep[$e.g.$ see][]{Fuentealba25}.  However, the next stage in our analyses will be to include environmental information (halo masses, nearest-neighbour density, pair/merger characteristics) in our analyses. This will both help us to include an understanding of satellite quenching at all stellar masses, that leads to the movement of galaxies in this plane, but also explore the impact of mergers in triggering AGN and leading to quenching events. This environmental analysis will form the basis of subsequent DEVILS papers.

\begin{figure*}
\begin{center}
\includegraphics[scale=0.55]{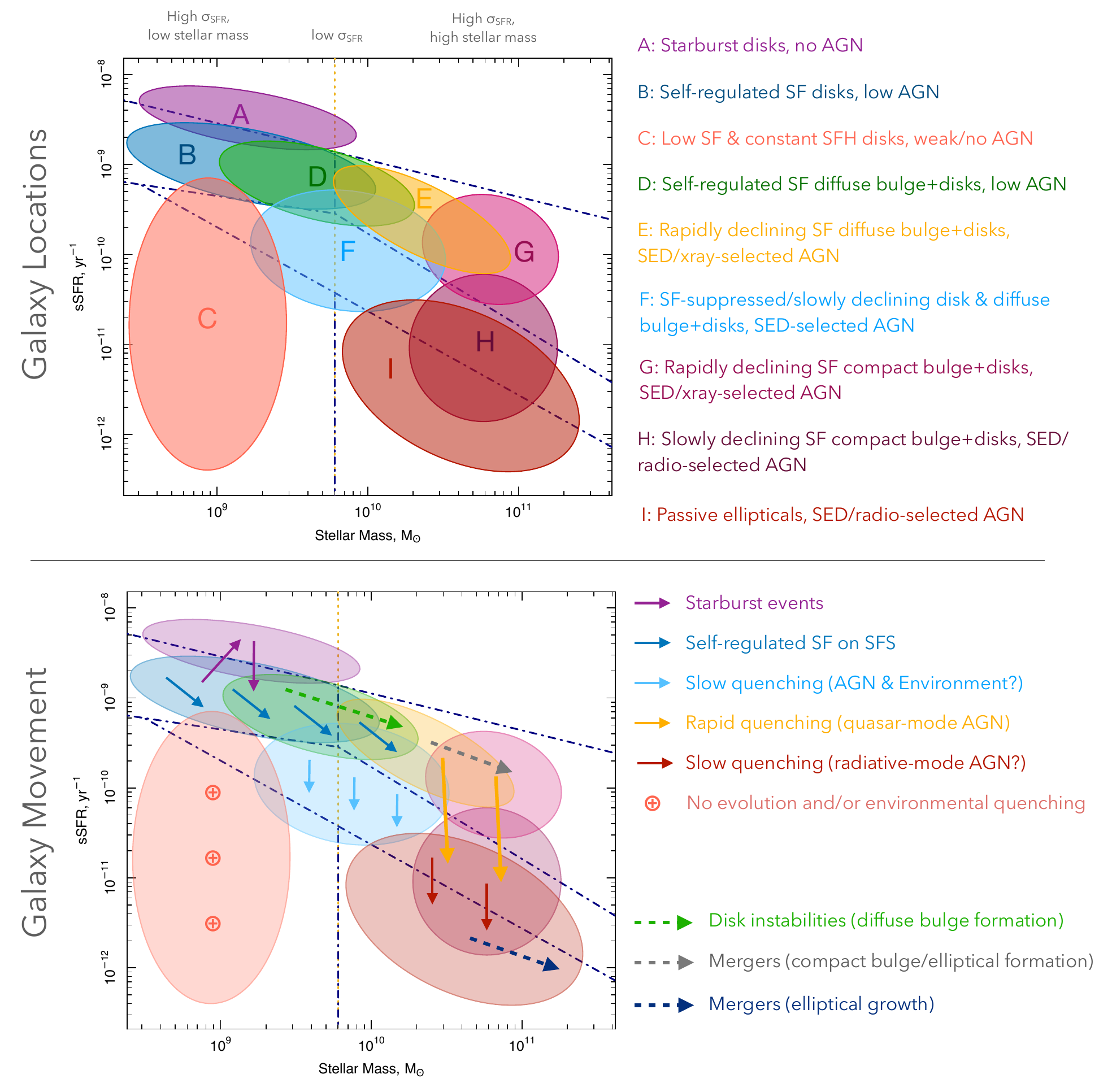}

\caption{Cartoon representation of the types of galaxies found in different regions of the sSFR-M$_{\star}$ plane (top) and how those populations are likely moving through the plane (bottom). See text for details.}
\label{fig:cartoon}
\end{center}
\end{figure*}

\subsection{Putting it all together}

In the final subsection of the paper we aim to put together a simplified picture of the key results outlined in both Paper I and II of this series, both in terms of the position of various subsamples of galaxies in the sSFR-M$_{\star}$ plane, and the average movement and cause of movement of those populations. In this work we have primarily explored the sSFR, M$_{\star}$, SFH, visual morphology, structure and AGN activity in our DEVILS galaxies. Using these properties we can loosely sub-divide the sSFR-M$_{\star}$ plane into overlapping regions containing galaxies with common properties. The top panel of Figure \ref{fig:cartoon} shows a cartoon representation of these regions, based on all of the previous analyses in this work. We have identified 9 different common types of galaxy, based on their combined properties and for each region we provide an estimation of the types of galaxies that reside there, in terms of their star-formation, morphology and AGN classification in the legend.  We also overlay the $0.4<z<0.55$ selection regions for common SFHs, described in paper I as the blue dot-dashed lines. We do not go into detail regarding the properties of these populations here (as they are described in Figure \ref{fig:cartoon}), but note how they lead to the various observed characteristics of the SFS and $\sigma_{SFR}$-M$_{\star}$ relation. We first see the tight SFS region (B, D, E, G) which changes in slope at M$^{*}_{\sigma-min}$, as galaxies transition from constant SFHs (B $\&$ D), to declining SFHs (E $\&$ G). We also see morphological trends as a function of stellar mass, with pure disks (B), diffuse bulge+disk (D $\&$ E) and compact bulge+disk (G) dominating the galaxy distribution at different stellar masses. At low stellar masses, we find two populations that deviate from the SFS leading to the increased SFR dispersion measurements; one from star-bursting disks (A) and another from low sSFR disks (C). At high stellar masses we also see a high SFR dispersion, driven by the quenching compact bulge+disk systems (H) and passive ellipticals (I). At intermediate stellar masses, we also separate the sSFR suppressed diffuse bulge+disk population (F), which show an increased number of \textsc{ProSpect}-select AGN and have slowly declining SFHs.

In the bottom panel of Figure \ref{fig:cartoon}, we then aim to show how galaxies in each population are moving as they evolve through the plane. We show solid arrows for the likely star-formation evolution of galaxies based on the evidence presented in this work, and also dashed arrows from a first-order approximation of the impact of structural changes and mergers that also move galaxies through the plane. Galaxies along the SFS at low stellar masses will evolve in a self-regulated fashion, gaining stellar mass with constant SFR (dark blue arrows), star-burst events (gas inflow, mergers) will push galaxies temporarily above the SFS, before they consume/expel the gas and drop back down (purple arrows). At the high stellar mass end, galaxies are rapidly dropping off the SFS due to radiative mode AGN feedback (orange arrows), kinetic mode AGN feedback causes slower declines in SFR in the most massive galaxies (red arrows). In the low stellar mass and low sSFR populations there is very little evolution as galaxies are not significantly growing or changing in SFH (red crosses), while at intermediate masses we see a slow decline in star-formation, likely caused by both AGN feedback and environmental quenching (light blue arrows). We also display additional arrows for morphological/structural evolution, where along the SFS galaxies likely grow diffuse/pseudo bulges through disk instabilities \citep[green dotted arrow, $e.g.$][]{Kormendy08} and minor mergers contribute to compact bulge growth \citep[grey dotted arrow, $e.g.$][]{Hopkins10}, and elliptical galaxies simply grow in stellar mass via both major mergers \citep{Taranu13} and minor mergers \citep{Bournaud07}, with likely little impact on their SFH. We note that this figure is qualitative and simply intended to provide a broad brush understanding of where different types of galaxies reside in this plane and to highlight the complexities in how this plane is evolving in time, both due to star-formation and merger processes. 

\textcolor{black}{Before concluding, we also highlight that the results and discussion in these papers, culminating in Figure \ref{fig:cartoon}, must be framed in the context of both the DEVILS survey sample selection, and our \textsc{ProSpect} methodology. Inherently, this biases our interpretation away from including rarer systems both due to sample size and cosmic variance (we only use the DEVILS D10 region that covers $\sim1.5$\,deg$^{2}$), and in terms of short-lived phases of galaxy evolution, such as rapidly fluctuating star-bursts, quenched then rejuvenated galaxies or galaxies that have recently finished a period of AGN activity but show little or no observational evidence of an AGN (which can not be easily identified or parameterised using \textsc{ProSpect}). Given our current analysis, we can say little about these systems, and as such, do not include them in Figure \ref{fig:cartoon}. However, it is likely that they would fill regions of this parameter space currently unoccupied by galaxies, such as rare star-bursting galaxies at high stellar masses (which are increasingly prevalent at earlier times). To progress and capture these rare and transitory sources we require similar samples, but with higher quality data and over much larger areas, and new approaches to better constrain galaxy SFHs. This is one of the key focuses of upcoming surveys such as the Wide Area VISTA Extragalactic Survey \citep[WAVES][]{Driver19}, which will provide similar samples to DEVILS but over $\sim50 \times$ the area used here, and new spectro-structural decomposition techniques using multi-wavelength high resolution imaging, to better constrain galaxy SFHs \citep[such as \textsc{ProFuse,}][]{Robotham22}.}

\section{Conclusions}

In paper I of this series we identified regions in the sSFR-M$_{\star}$ plane that contain galaxies with similar recent SFH shapes.  We then displayed how the movement of these different populations through the plane leads to the overall evolution of the SFS and the SFR dispersion as a function of stellar mass ($\sigma_{SFR}$-M$_{\star}$). Here we explore the physical characteristics of galaxies with common SFH shapes. We aim to identify the physical drivers of changes to their SFR, which then drive them away from the tight locus of the SFS, and ultimately lead to the high SFR dispersion populations identified by the $\sigma_{SFR}$-M$_{\star}$ relation. With this goal in mind we specifically focus on two regimes: the high SFR dispersion population at high stellar masses (which are associated with rapidly declining SFHs) and the high SFR dispersion population at low stellar masses. 

At the high stellar mass end, we find that the high SFR dispersion population (with rapidly declining SFHs) is strongly associated with an overabundance of AGN. We also find that this overabundance of AGN evolves with the high SFR dispersion population across redshift. This suggest that AGN feedback is the likely cause of the rapidly declining SFHs, which in turn leads to the high SFR dispersion population. This is consistent with the simulation predictions of  \cite{Katsianis19}, and others.    

At the low stellar mass end we find that galaxies display a mix of both rapidly increasing SFHs at high sSFRs, and constant SFHs that have remained similar for the entirety of the galaxy's history. Modulo the fact that our \textsc{ProSpect} analysis can not capture short timescale variations in SFH, this suggests that the high SFR dispersion population at low stellar masses is produced by a combination of stochastic star-formation processes and galaxies with a large variation in normalisation of constant SFHs.    

In combination these results provide an observational constraint on the physical properties that are driving the distribution and evolution of galaxies within the sSFR-M$_{\star}$ plane. The final missing piece of this picture is the environmental impact on the star-forming properties of galaxies (which is not included here). This will form the basis of subsequent papers using the DEVILS sample.

\section*{Acknowledgements}

LJMD, ASGR, and SB acknowledge support from the Australian Research Councils Future Fellowship scheme (FT200100055 and
FT200100375). JET was supported
by the Australian Government Research Training Program (RTP)
Scholarship. MS acknowledge support from the Polish National Agency for Academic Exchange (Bekker grant BPN/BEK/2021/1/00298/DEC/1), the European Union's Horizon 2020 Research and Innovation programme under the Maria Sklodowska-Curie grant agreement (No. 754510). Parts of this research were
conducted by the Australian Research Council Centre of Excellence
for All Sky Astrophysics in 3 Dimensions (ASTRO 3D), through
project number CE170100013. DEVILS is an Australian project
based around a spectroscopic campaign using the Anglo-Australian
Telescope. DEVILS is part funded via Discovery Programs by the
Australian Research Council and the participating institutions. The
DEVILS website is \url{devils.research.org.au}. The DEVILS data are
hosted and provided by AAO Data Central (\url{datacentral.org.au}).

\section{Data Availability}

Data products used in this paper are taken from the internal DEVILS
team data release and presented in \cite{Davies21} and \cite{Thorne21}. These catalogues will be made public as part of the DEVILS
first data release described in Davies et al. (in preparation).

\appendix

\section{Structure and Morphology}  
\label{sec:Morph}

\textcolor{black}{In this Appendix, we discuss the various lines of exploration to determine the potential correlation between galaxy structure/morphology and common SFHs. We ultimately find that there is no strong correlation between the morphological or structural parameters we explore and recent SFH, as such we only discuss the key results in the main body of the paper.}   

\begin{figure*}
\begin{center}
\includegraphics[scale=0.37]{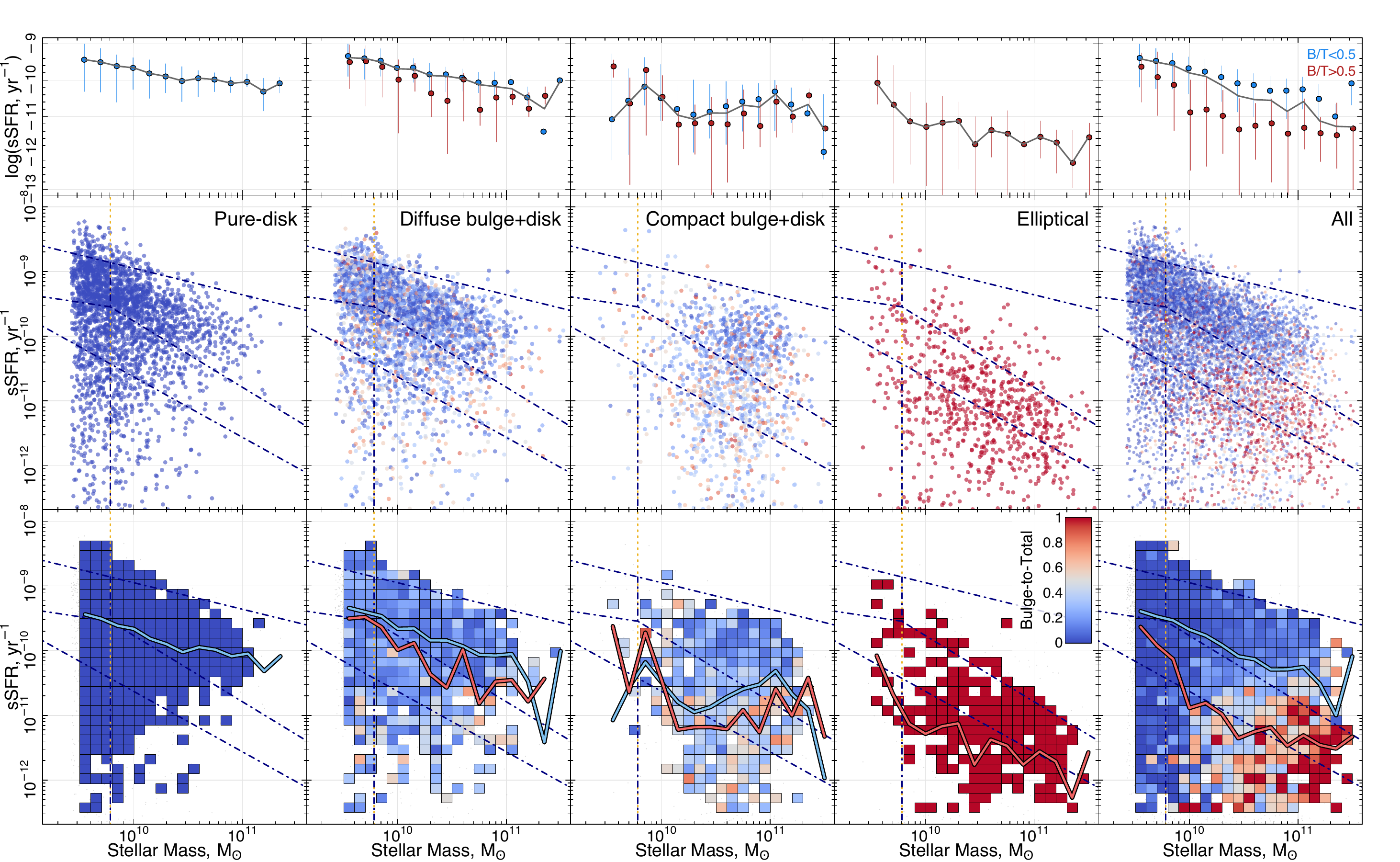}

\caption{The distribution of visual morphology (columns) and bulge-to-total measurements (colours) across the sSFR-M$_{\star}$ plane at $0.4<z<0.7$ (note we combine two of our redshift bins to increase sample size, but over-plot selection regions at $0.4<z<0.55$ for reference). In the middle row we show individual galaxies, while in the bottom row we show a binned version coloured by the median B/T value in each bin.  In the top row we show the running median (points) and interquartile range (bars) of sSFRs for the total (grey), B/T$<$0.5 (blue) and B/T$>$0.5 (red) samples - which are also over-plotted on the bottom row. Note that visual morphology classes of `disk' and `elliptical' are assigned a B/T of 0 and 1 respectively - and as such there is no colour variation. We find that, as expected, different morphological classes occupy different regions of the sSFR-M$_{\star}$ plane, with disk and diffuse bulge+disk systems predominantly falling on the traditional SFS at lower stellar masses, compact bulge+disk systems sitting at the high stellar mass end of the traditional SFS region, and extending down through the `green valley' to the passive region, and ellipticals falling in the `green valley' and passive regions at high stellar masses. We also see weak trends with bulge-to-total, where for two-component systems, objects appear to be become more bulge dominated as you fall below the SFS.  }

\label{fig:BTplane}
\end{center}
\end{figure*}

\subsection{Visual Morphology}  

To explore the morphological/structural distribution of the galaxies in each of our regions, we first use the visual morphological classifications of \cite{Hashemizadeh21}. In that work the HST images of M$_{\star}>$10$^{9.5}$M$_{\odot}$ DEVILS galaxies were visually classified into pure disk, diffuse (pseudo) bulge+disk, compact (classical) bulge+disk, elliptical, and a number of categories of unclear classifications (hard/merger, too compact, too faint, stars, unsure). Here we do not show the `unclear' classifications as they only make up a very small fraction of the sample ($<0.1\%$). Note that this stellar masses limit is imposed as below this even HST resolution data cannot be used to visual classify galaxies in the DEVILS sample due to faintness and/or small angular sizes. 

Figure \ref{fig:BTplane} first displays the distribution of different morphological classes across the sSFR-M$_{\star}$ plane (columns) at $0.4<z<0.7$ (we combine two of our redshift bins to increase the number of sources displayed). Here we see clear trends of different morphological types occupying different regions of this parameter space. Pure disk systems fall largely on the SFS at lower stellar masses, diffuse bulge+disk systems also lie on the SFS and are more preferentially found at higher stellar masses, compact bulge+disk systems are found at higher stellar masses  and span the SFS region to the passive region over a broad range in sSFRs, while the elliptical galaxies fall predominately in the slow quenching/passive regions. These results are not surprising, and are consistent with many other previous studies exploring the distribution of morphology with star-formation and stellar mass \citep[$e.g.$][]{Schawinski14, Otter20, Cook19}.         

To take this further, and explore how morphology varies over our new selection regions, the left column of Figure \ref{fig:regionComp} displays the fraction of galaxies in each visual class in each region. Here we once again see a number of well known trends: galaxies on the SFS and in the star formation increasing regions are dominated by pure disk and diffuse bulge+disk galaxies, while high stellar mass passive galaxies are dominated by elliptical and compact bulge+disk galaxies. Interestingly the `rapid quenching' galaxies, while showing a spread of morphological types, are more skewed towards pure disk and diffuse bulge+disk morphologies, similar to the SFS galaxies. Whereas the `slow quenching' systems are a mixed bag of different morphological types, with less skew to disk-like systems. This potentially suggests that these galaxies show morphological evolution that is coincident with the quenching event. For example, if one assumes (to first order) a quenching sequence of SFS $>$ Rapid Quenching $>$ Slow Quenching $>$ passive,  this can also be followed as a change from pure disk $>$ diffuse bulge+disk $>$ compact bulge+disk $>$ Elliptical. We note that this trend is also observed when using the $\Delta$SFH$_{200\mathrm{\,Myr}}$ values directly (Figure \ref{fig:regionComp}). However, this is very qualitative in terms of any evolutionary sequence, and says nothing about whether or not the decline in star formation and morphological change are causally linked.  Potentially even more interesting is that the `low mass passive' systems show distinctly different morphological types to the high mass passive systems, being strongly dominated by disks. This could potentially suggest two separate quenching routes at different stellar masses, one that changes morphology and one that does not  \citep[$e.g.$ see][who also suggest two quenching pathways, one fast and one slow]{Schawinski14}. This is also seen in Figure \ref{fig:BTplane}, where the lower mass passive sources are dominated by pure disk and diffuse bulge+disk morphologies. However, with visual morphological information alone it is difficult to assess if the morphological change is driving the movement of galaxies through the sSFR-M$_{\star}$ plane, is caused by it, or is serendipitously occurring at the same time.

\begin{figure*}
\begin{center}
\includegraphics[scale=0.45]{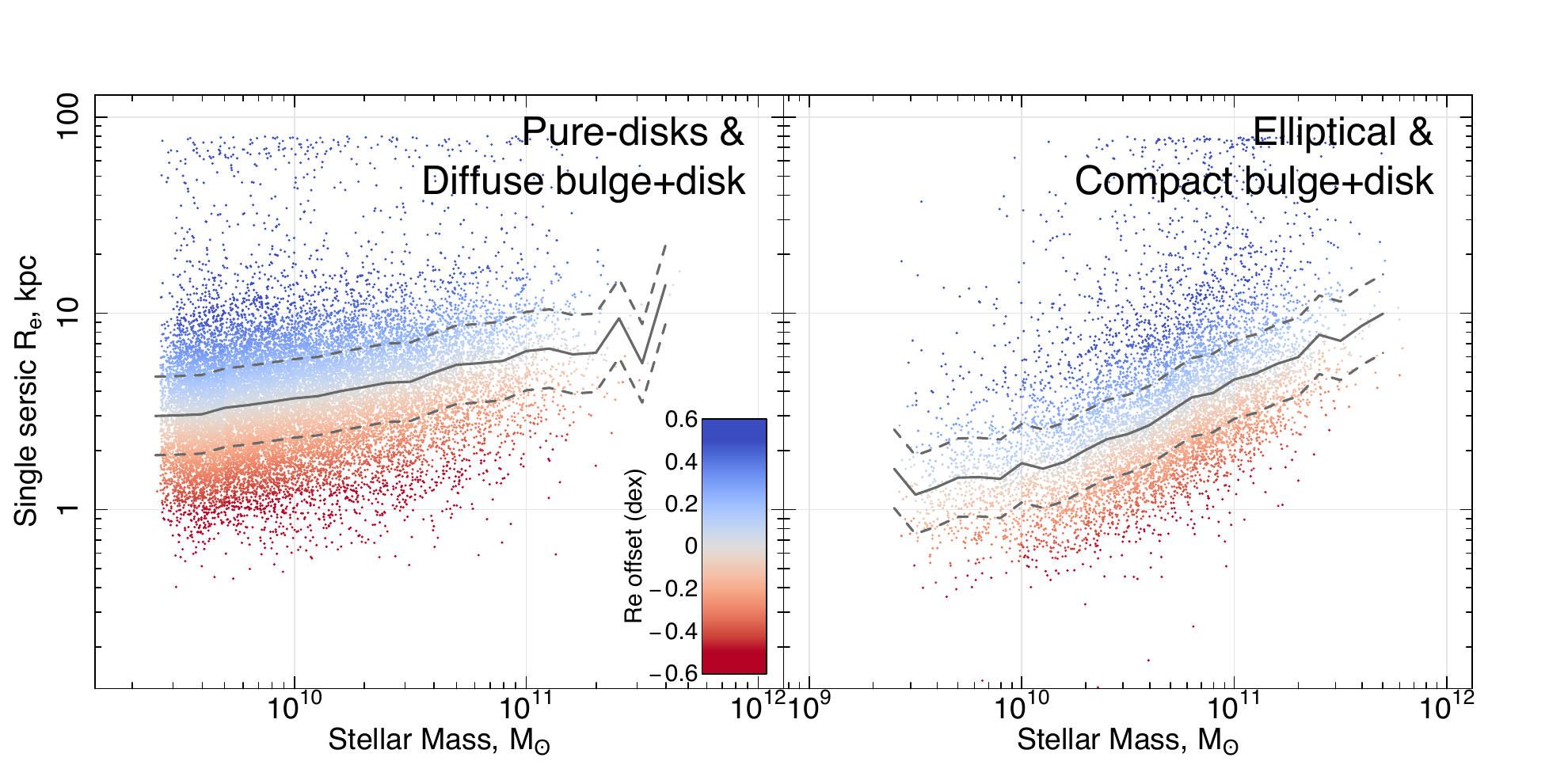}
\includegraphics[scale=0.37]{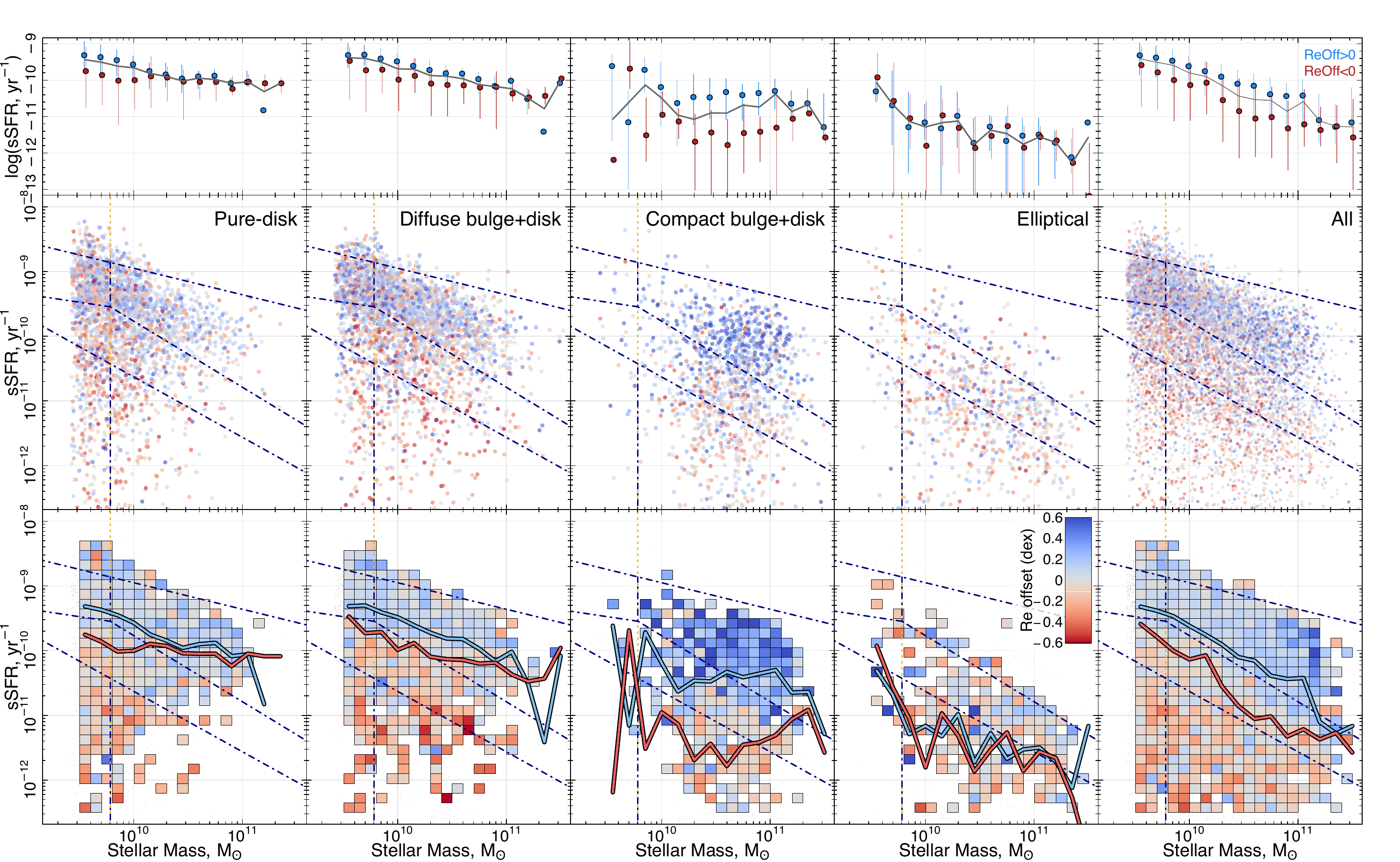}

\caption{Top: The measure of `compactness' of a galaxy with respect to the `typical' galaxy at a given stellar mass and visual morphology. We split galaxies into two different visual morphological classes: pure disks \& diffuse bulges+disk, and ellipticals \& compact bulges+disks. For each class we plot the stellar mass against single S\'{e}rsic effective radius. In bins of $\Delta$log$_{10}(M$$_{\odot}$)=0.1 we measure the median S\'{e}rsic effective radius (grey solid lines). We then calculate the R$_{e}$ offset as the distance from the median S\'{e}rsic effective radius at that stellar mass (colour of points) in dex, such that R$_{e}$ offset=0 lies on the grey solid line. The dashed grey lines show an R$_{e}$ offset of $\pm$0.2\,dex. Bottom: The distribution of compactness (R$_{e}$ offset) across the sSFR-M$_{\star}$ plane at $0.4<z<0.7$ and split by morphological class, as in Figure \ref{fig:BTplane}. In the middle panels we show individual galaxies, while in the bottom panels we show a binned version coloured by the median R$_{e}$ offset value in each bin.  In the top row we show the running median (points) and interquartile range (bars) of sSFRs for the total (grey), R$_{e}$ offset$>$0 (blue) and R$_{e}$ offset$<$0  (red) samples - which are also over-plotted on the bottom row. We see a clear trend of compactness with position relative to the SFS, with galaxies being more compact to lower sSFR, irrespective of morphology.   }

\label{fig:ReSel}
\end{center}
\end{figure*}

\subsection{Bulge-to-Total Ratio}

Next, we consider the difference in light-weighted bulge-to-total (B/T) measurements from the structural decomposition analysis of DEVILS galaxies presented in Cook et al (in prep). Briefly, this work used the \textsc{ProFit} \citep{Robotham17} and \textsc{ProFuse} \citep{Robotham22} packages to decompose the HST imaging for DEVILS galaxies in the D10 region into bulge and disk components. Here we take the HST F814W light-weighted bulge-to-total (B/T) values from Cook et al.  We use the visual morphological classifications of \cite{Hashemizadeh21} and assign objects which are visually classified as pure disks a B/T=0 and objects which are visually classified as ellipticals a B/T=1. For all other morphological classes we use the B/T values directly. Note that this analysis is once again only performed on M$_{\star}>$10$^{9.5}$M$_{\odot}$ galaxies, and as such we do not have B/T values for galaxies below this limit. 

Figure \ref{fig:BTplane} also displays the distribution of B/T values across the sSFR-M$_{\star}$ plane and by morphological type. For both pure disk and elliptical morphologies we assign the B/T value, so no further information is displayed. For the 2-component systems, we still see a weak trend of B/T increasing to lower sSFR at a fixed stellar mass and visual morphology, particularly for compact bulge + disk systems \citep[$e.g.$ similarly to][]{Cook19}. 

The middle column of Figure \ref{fig:regionComp} shows the B/T distribution in each of our regions, with fraction of galaxies at very low and high B/T given in the legend. Firstly, as you move below the SFS the fraction of galaxies with B/T$>0.9$ increases significantly. This is a well known trend of galaxies being more dominated by elliptical-like morphologies in the passive region of the sSFR-M$_{\star}$ plane. However, this is also true for the `low mass passive' galaxies, which do show distinctly higher numbers of B/T$>0.9$ systems than galaxies with constant SFHs on the SFS at the same stellar mass. However, we note again here that the Cook et al work only extends to M$_{\star}>$10$^{9.5}$M$_{\odot}$ galaxies, and thus only contains the most massive galaxies in the `low mass passive' region. So it is potentially not surprising that they share structural characteristics with the high stellar mass passive population (albeit note the morphological differences highlighted above).

Galaxies on the SFS and in the SF increasing region show very similar B/T distributions and are both dominated by pure disk galaxies, as expected. The most interesting result of this analysis is that the `rapid quenching' population shows very similar B/T distributions to the SFS and the SF increasing galaxies, whereas the `slow quenching' galaxies show much higher B/T$>0.9$ fractions - consistent with their visual morphological assignments. Once again, if this is an evolutionary sequence (galaxies rapidly quenching and then slowly quenching), it suggests that despite the fact that `rapid quenching' galaxies are likely going through a rapid quenching event, their structural characteristics are still very similar to galaxies on the SFS, but then change as they decline further in sSFR. This potentially indicates that galaxies do not significantly change structure prior to their quenching event, but do so during or after the quenching, $i.e.$ a change in structure/morphology does not trigger the quenching event \citep[$e.g.$ see][]{Cortese19}.

\subsection{Compactness}  

Finally in terms of structural characteristics, we look to another metric which has previously been linked to galaxies moving through the sSFR-M$_{\star}$ plane, the compactness of galaxy sizes in comparison to typical galaxies at the same stellar mass \citep[$e.g.$][]{Faisst17, Wang18, Zolotov15}. To obtain a measure of the compactness of our galaxies, we use the effective radius (Re) for single S\'{e}rsic fits to the DEVILS galaxies produced in Cook et al (in prep). We then combine our sample into pure disk and diffuse bulge+disk, and compact bulge+disk and elliptical morphologies. Cook et al  show that these morphological types are consistent in the size-mass relation and that the slope and normalisation of the size-mass relations for these types evolve little with redshift. The top panels of Figure \ref{fig:ReSel} shows the size-mass relation for these two different morphological selections. We then measure the median Re for each morphological selection in $\Delta$log$_{10}$(M$_{\star}$)=0.1 bins and determine the R$_{e}$ offset (in dex) of all galaxies from the median, which we call `R$_{e}$ offset'. The points in Figure \ref{fig:ReSel} are colour coded by this offset. This metric essentially gives a measure of how compact a galaxy is with respect to all other galaxies of the same morphology and stellar mass, where a positive R$_{e}$ offset is a galaxy that is atypically diffuse, while a negative R$_{e}$ offset galaxy is atypically compact.

The bottom panels of Figure \ref{fig:ReSel} show the distribution of R$_{e}$ offset values across the sSFR-M$_{\star}$ plane, split by the four morphological classes described above. Interestingly we find that for all morphological classes there is a relatively strong trend of compactness increasing as galaxies move below the SFS, and more compact galaxies having lower median sSFR (particularly for compact bulge+disk systems). This suggests that as galaxies decline in sSFR away from the SFS their \textit{measured} structural characteristic change, such that their light profile become more compact. 

The initial  explanation for this may be that the quenching of galaxies and their compactness are linked. However, a similar (and potentially more likely) effect can be obtained via disk fading \citep[$e.g.$][]{Head14, Bremer18} leading to an increase in measured compactness with no real change in true structural characteristics \citep[however, c.f.][on the ability of disk fading to account for changes in morphological type]{Christlein04}. In this scenario, galaxy disks fade as star-formation declines while non-star-forming bulges remain constant, leading to more centrally concentrated light profiles \citep[$e.g.$ see][]{Croom21}. This is exactly the trend we see in Figure \ref{fig:ReSel}. In addition, the fact that the strongest change in compactness with sSFR occurs in compact bulge+disk systems, and these are likely the sources what would be most strongly affected by disk fading (prominent compact bulge and extended disk), adds weight to this possibility. 

Moreover, specifically for this work we are aiming to explore how \textit{change} in SFH correlates with galaxy properties. In Figure \ref{fig:ReSel} we find that galaxies in the region which contains predominantly rapidly declining SFHs are actually the most diffuse systems for their stellar mass at all morphologies. Following this the right column of Figure \ref{fig:regionComp} then shows the distribution of R$_{e}$ offset values for each of our SFH selection regions in more detail. We find that galaxies on or above the traditional SFS region have slightly higher R$_{e}$ offset values (more diffuse) than the typical galaxy, while passive galaxies have significantly smaller R$_{e}$ offset values (more compact) than the typical galaxy (as is noted above). However, most notably the rapidly quenching galaxies have a very similar compactness distribution to the SFS population. This once again suggests that there is no clear structural difference between galaxies which are in the initial stages of quenching, and those which are not - suggesting the quenching process is not preceded by a structural change.

\section{AGN}
\label{sec:AGNapp}

\textcolor{black}{In this Appendix, we discuss the details of the comparison between SFR, stellar mass, morphology and AGN (selected using different methodologies). Here we provide a detailed comparison of this complex multi-dimensional parameter space for completeness and to aid the reader in interpreting the results presented in this work. However, we simply summarise the key findings in the main body of the paper.}

\begin{figure*}
\begin{center}
\includegraphics[scale=0.37]{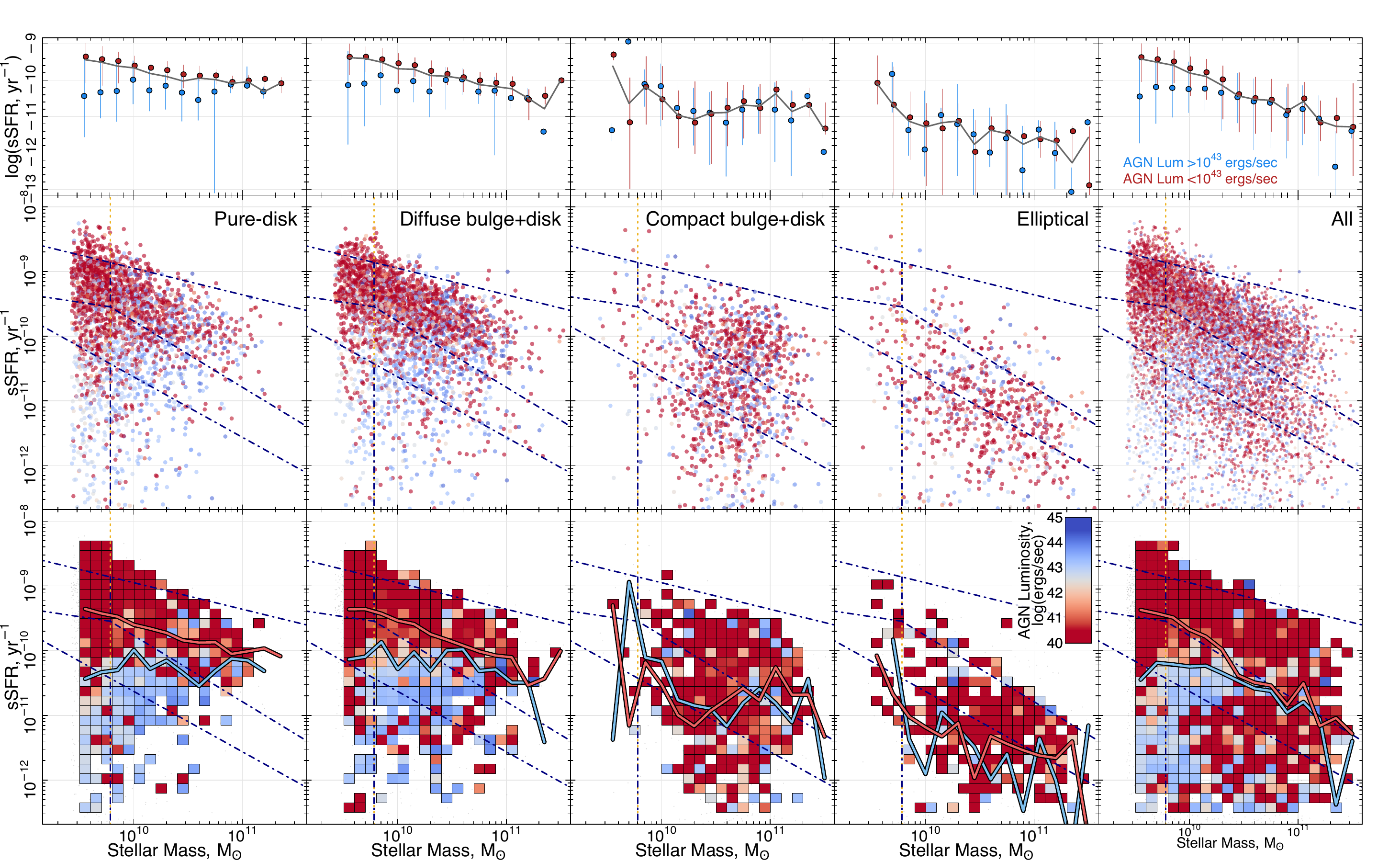}

\caption{The distribution of visual morphology (columns) and \textsc{ProSpect} AGN luminosity measurements (colours) across the sSFR-M$_{\star}$ plane at $0.4<z<0.7$, as in Figure \ref{fig:BTplane}. In the middle row we show individual galaxies, while in the bottom panels we show a binned version coloured by the median AGN luminosity value in each bin. In the top row we show the running median (points) and interquartile range (bars) of sSFRs for the total (grey), AGN luminosity $>10^{43}$ergs\,s$^{-1}$ (blue) and AGN luminosity $<10^{43}$ergs\,s$^{-1}$ (red) samples - which are also over-plotted on the bottom row. This range essentially selects for systems that have a robust AGN (blue) and a non-robust or weak AGN (red), based on the AGN completeness limits of \citet{DSilva23}. We see that for pure disk and diffuse bulge+disk systems there is an over-density of luminous AGN below the SFS and that sources with bright AGN have systematically lower sSFRs. }

\label{fig:ANGplane}
\end{center}
\end{figure*}

\begin{figure}
\begin{center}
\includegraphics[scale=0.6]{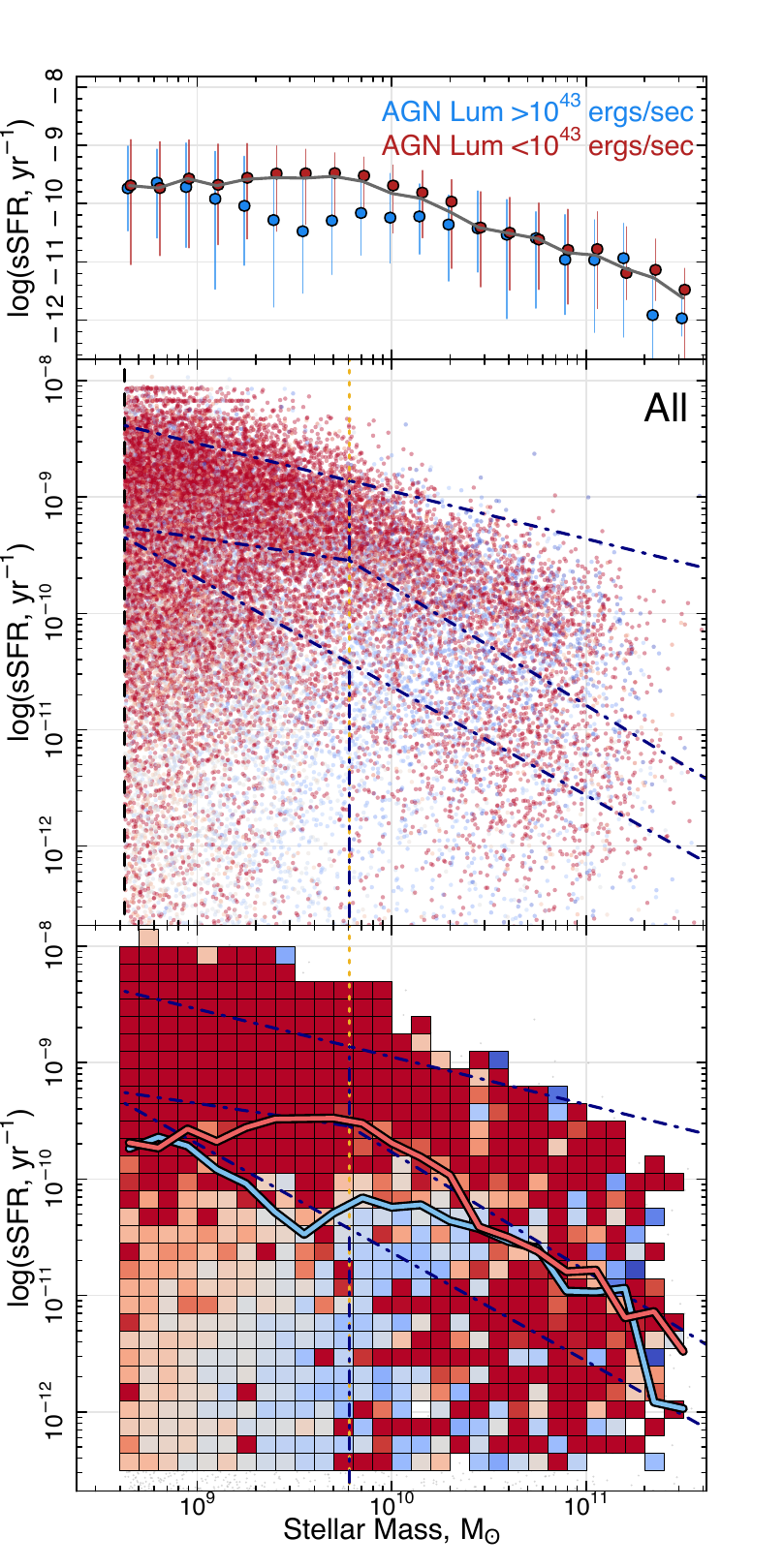}

\caption{The same as the rightmost panels of Figure \ref{fig:ANGplane}, but extending down to the stellar mass completeness limit at $0.4<z<0.7$ (not just for galaxies that have morphological classifications). Here we see that the incidence of luminous AGN and the impact of reduced median sSFRs in sources with luminous AGN, only strongly occurs in $10^{9}<\mathrm{M}_{\star}/\mathrm{M}_{\odot}<10^{10.25}$ galaxies. As we extend to lower stellar masses the effect is removed. This suggests that \textsc{ProSpect}-selected AGN may only be significantly impacting the star formation in galaxies at a particular stellar mass.}

\label{fig:ANGplaneAllex}
\end{center}
\end{figure}

\subsection{AGN selected via \textsc{ProSpect} MIR flux fraction}  
 \label{sec:AGNfrac}

First, we select potential AGN as galaxies that have a significant fraction of their bolometric luminosity arising from the AGN component in their \textsc{ProSpect} SED fit.  To calculate this, \cite{Thorne22} use \textsc{ProSpect} to determine the fraction of bolometric light that can be attributed to an AGN at mid-infrared wavelengths (AGN fraction). See \cite{Thorne22} for a full description of this process. However, they then suggest that an AGN fraction $>0.1$ is indicative of the presence of an AGN within the host galaxy, and use this selection in subsequent analysis. 

In the left column of Figure \ref{fig:regionCompAGN}, we show the distribution of AGN fraction for galaxies in each of our sub-regions of the sSFR-M$_{\star}$ plane as dotted lines. We also note the percentage of galaxies in each region that have AGN fraction $>0.1$ in the top left legend. We remind the reader that a similar panel for sources selected on $\Delta$SFH$_{200\mathrm{\,Myr}}$ directly is also given in Figure \ref{fig:regionCompAGN}. We find that passive galaxies (both at low and high stellar masses) have high fractions of galaxies with AGN fraction $>0.1$, with low mass passive systems showing many galaxies that have their light distribution dominated by the AGN. Conversely, SFS and star-formation increasing galaxies have low fractions of galaxies with AGN fraction $>0.1$, with the majority of systems containing no (or a very sub-dominant) AGN. The slow quenching galaxies sit between these with many low-to-intermediate AGN fraction systems, but less dominant AGN than the passive galaxies. Interestingly, the rapidly quenching galaxies also appear very similar to the SFS galaxies in terms of their AGN fraction distribution and overall percentage of sources with AGN fraction $>0.1$, suggesting that in terms of the bolometric fraction of light attributed to an AGN, these samples are largely the same.    

One important caveat to this, noted in \cite{Thorne22}, is that it is difficult to constrain the AGN contribution to a galaxy's bolometric light distribution without a FIR detection (which are the shallowest imaging bands in the DEVILS sample) - also see \cite{Kirkpatrick23}. As such, \cite{Thorne22} also include a sub-sample of galaxies that have a detection in at least one of the FIR bands, that they suggest have more robust AGN fraction measurements. In the left column of Figure \ref{fig:regionCompAGN} we also show the AGN fraction distribution for the galaxies which are detected in the FIR as solid lines, and percentage of galaxies in each region that have a FIR detection and AGN fraction $>0.1$ in the top right legend. Including this extra constraint only significantly impacts the distribution of lower stellar mass galaxies (SF increasing and low mass passive samples), but does not significantly affect the overall trends in the results. As such, these trends are robust to the measurement of AGN fraction.                    

\subsubsection{Low stellar mass AGN?}

One notable trend in the above is that the `Low mass passive' population has a very large percentage of galaxies with high AGN fraction, which is in fact even higher than for the high stellar mass passive population. This is true for the full population and just considering objects with a FIR detection (we note here that FIR-detected objects in the low mass passive regime do not occupy a specific region of the sSFR-M$_{\star}$ plane). Finding such a large incidence of low-stellar mass AGN one would immediately think that this is an artefact of the fitting process. However, when inspecting the SED-fit outputs for these objects, the majority do have a MIR excess that is difficult to explain using a stellar+dust model alone ($i.e.$ they require the inclusion of an AGN component), particularly those with FIR detections. In addition, there are models that predict effective feedback from AGN in low stellar mass galaxies \citep[$e.g.$][]{Dashyan18}. 

However, as this result is surprising, we also explore if these objects could be placed erroneously in the sSFR-M$_{\star}$ plane as passive systems due to measurement and/or fitting error. First, we consider that a lack of robust photometric redshift measurements at the low stellar mass end could potentially mean that these objects are in fact much larger galaxies at higher redshift. We find that just 5\% of this population have a spectroscopic redshift, but in 90\% of those cases the spectroscopic redshift agrees with the photometric redshift ($\Delta z<0.1$). In cases where there is no spectroscopic redshift, the photometric redshifts do not have a significantly large error, and inspection of the SEDs compared to the best fit photometric redshift suggests that they are largely correct in most cases.  Next we explore if the potentially erroneous fitting of an AGN component to these systems significantly changes their measured SFR and stellar mass, such that they would no longer fall in the same region of the sSFR-M$_{\star}$ plane. In \cite{Thorne21} they also provide SFR and stellar mass measurements from \textsc{ProSpect} \textit{without} fitting an AGN component. Hence, we compare directly the position of this sample in the sSFR-M$_{\star}$ plane when fitted with and without an AGN component, and while we find that (as expected) the sSFRs of these sources are larger when fitting without an AGN model, the majority would still remain within the `low mass passive' selection window, and only a small fraction would then reside on the SFS \citep[also see][]{Zou22}. We also note that in our original D22 analysis of the $\sigma_{SFR}$-M$_{\star}$ relation we used the non-AGN fits of \cite{Thorne21}, and hence the potential erroneous sSFR measurements from fitting an AGN component do not drive the high SFR dispersion seen at these low stellar masses. So in summary, we can not rule out that this population of low mass passive systems have biased sSFR measurements due to the erroneous inclusion of an AGN component, but this will likely not significantly effect the conclusions drawn in this paper.

Following this, the likely high incidence of large AGN fraction values in passive systems,  potentially suggests that there is a correlation between bolometric AGN fraction and sSFRs across all stellar masses. However, AGN fraction says little about the overall AGN emission (and likely feedback processes) as it is relative to the overall light distribution ($i.e.$ a very low sSFR and even weak AGN could have a high AGN fraction value).  

\subsection{AGN selected via \textsc{ProSpect} AGN luminosity}  
 \label{sec:AGNLum}

To better constrain the impact of AGN on the star-formation in host galaxies, we then also explore the bolometric luminosity of the \textsc{ProSpect} AGN component across the sSFR-M$_{\star}$ plane. The analysis of \cite{Thorne22} provides these measurements for all galaxies in our sample. In Figure \ref{fig:ANGplane} we show the distribution of AGN bolometric luminosities across the plane, split by visual morphology and showing the full sample in the right-hand panel. \textcolor{black}{We opt to split by visual morphology here as previously it has been noted there are strong trends between visual morphology and stellar mass, SFR and AGN fraction. As such, we wish to explore if there are any trends across the sSFR-M$_{\star}$ plane which become apparent when isolating for morphology}.This figure is colour-coded in such a way, that essentially red indicates no secure AGN and blue indicates the presence of an AGN \citep[based on ths  DEVILS AGN detection limits of][]{DSilva23}. First, we see that for compact bulge+disk and elliptical morphologies, there is no significant trend between position within the plane and AGN luminosity. Galaxies show a broad range of different AGN luminosities, that span all sSFRs and stellar masses. There is also no difference in the median sSFRs for different AGN luminosity ranges (top row). However, for pure disk and diffuse bulge+disk morphologies, we do see a clear difference between galaxies hosting bright AGN and those which do not. In all but the highest stellar mass systems, we see a large fraction of AGN in systems sitting below the SFS in our `slowly quenching' and passive regions. The top row of Figure \ref{fig:ANGplane} also shows that the median sSFRs of galaxies hosting luminous AGN are systematically lower than those not hosting luminous AGN. This potentially indicates that in disk and diffuse bulge+disk galaxies, the presence of a luminous AGN has suppressed star formation by $\sim0.5-1$\,dex. 

However, Figure \ref{fig:ANGplane} only shows sources for which we have visual morphological classifications. Taking this further, in Figure \ref{fig:ANGplaneAllex} we also show the full sSFR-M$_{\star}$ plane, for all morphological types and down to the stellar mass completeness of our sample. Here, we see that the impact of these luminous AGN in reducing the median sSFR of galaxies, is restricted to stellar masses around $10^{9}<\mathrm{M}_{\star}/\mathrm{M}_{\odot}<10^{10.25}$, and at lower/higher stellar masses, we do not see any significant effect. This is intriguing and potentially suggests the SED-selected AGN are only having a strong impact on galaxies at a particular stellar mass range. However, we also note here that through the \textsc{ProSpect} analysis we are identifying sources based on `instantaneous' AGN luminosity, so this does not say that AGN caused these galaxies to drop below the SFS, but does not rule out that they are still in the process of declining.

\section{Long timescale SFHs of low stellar mass galaxies}                                       
\label{sec:starburst}

\begin{figure*}
\begin{center}
\includegraphics[scale=0.16]{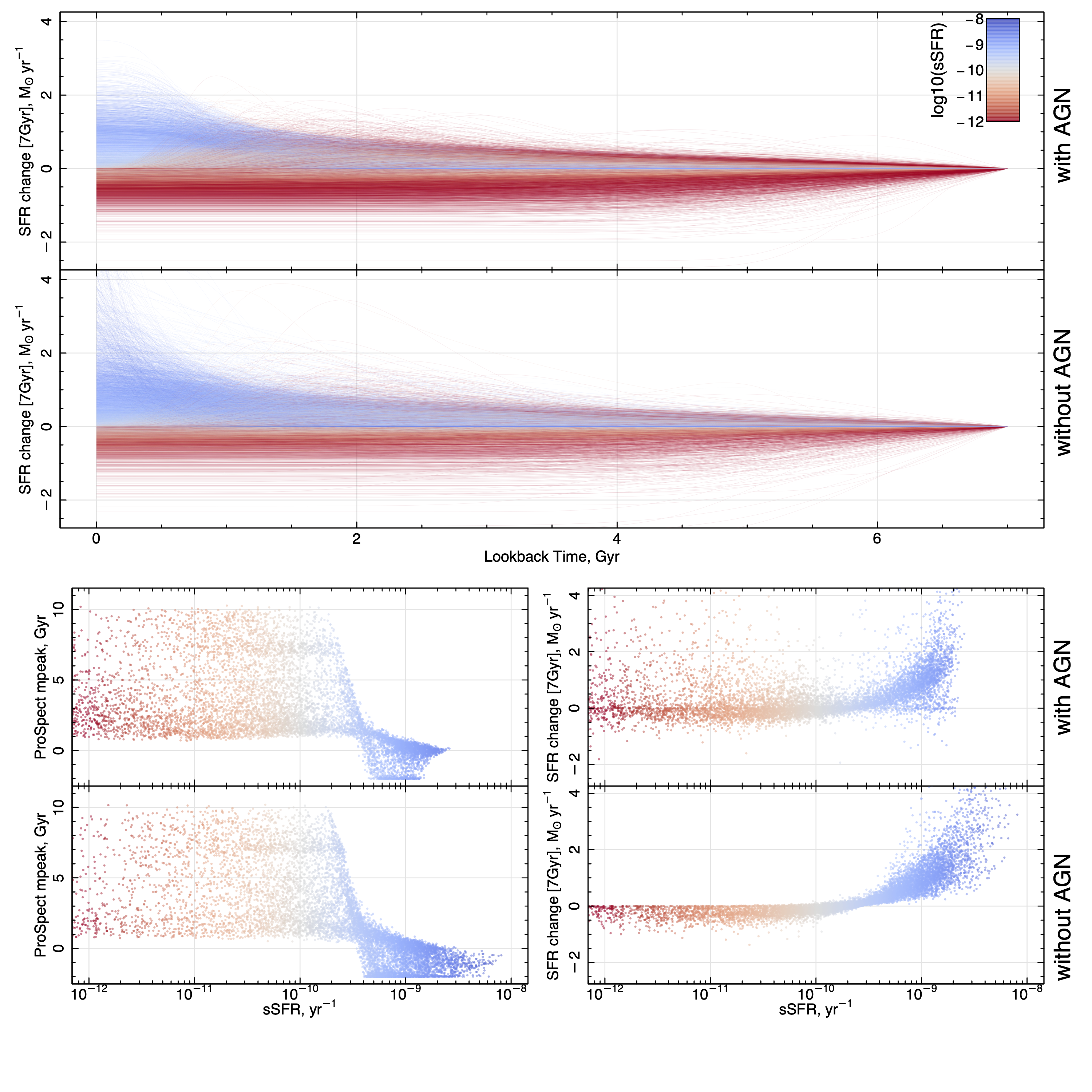}

\caption{The long-timescale SFH changes for galaxies at $0.25<z<0.7$ with 8.8$<$log$_{10}$(M$_{\star}$/M$_{\odot}$)$<$9.2 and constant $\Delta$SFH$_{200\mathrm{\,Myr}}$. Top panels show the change in SFR over the last 7\,Gyr colour-coded by $t=0$ sSFR for both \textsc{ProSpect} fits with (top) and without (bottom) and AGN component. The bottom rows show the \textsc{ProSpect}-mpeak (peak look-back time in SFH) and final SFR change over 7\,Gyr.  The sample splits into two populations divided at sSFR$\sim10^{-9.5}$\,yr$^{-1}$. One which sits on the SFS and formed much later, and one which sits in the `low mass passive' regime, which formed early and has had relatively constant SFR over the last 10\,Gyr. These do appear to be a truely old, low mass passive population based off their \textsc{ProSpect} fits. }
\label{fig:longSFH}
\end{center}
\end{figure*}

\textcolor{black}{In this appendix we further explore the possibility that the large SFR scatter at the low stellar mass end of the sSFR-M$^{\star}$ plane is caused by a combination of stochastic SFHs $and$ long duration constant SFHs.  Assuming that this result is not a limitation of our fitting process, it is worth asking the question:} at a fixed stellar mass and for galaxies with constant SFH, what potentially governs the normalisation of the SFH leading to a large diversity in current sSFR? To explore this further we select galaxies at $0.25<z<0.7$ with 8.8$<$log$_{10}$(M$_{\star}$/M$_{\odot}$)$<$9.2 and constant $\Delta$SFH$_{200\mathrm{\,Myr}}$ as a representative sample of low stellar mass, flat SFH sources.  Figure \ref{fig:longSFH} displays the long time-scale (7\,Gyr, $\sim$age of Universe at $z=0.7$) SFH differences for this sample colour-coded by current sSFR, where red colours have lower sSFRs and blue colours have higher sSFRs. For comparison we also show \textsc{ProSpect} SFH fitted with a possible AGN component (top row) and without (bottom row). We also show the \textsc{ProSpect-mpeak} value in Gyr (the peak in the galaxy's SFH, which is roughly the mean stellar age) and the final SFR change over the last 7\,Gyr below.  

Firstly, we note that while we see differences between the SFH change in the \textsc{ProSpect} runs with possible AGN and not (particularly for the higher sSFR galaxies), the overall trends particularly in mpeak and total SFR change are \textcolor{black}{similar. For example, they both have transition points at the same sSFR, and the same trends with sSFR}. Hence, these differences are not likely to be strongly driven by the inclusion of AGN in the fitting process. Following this, we see distinct differences in the populations as a function of sSFR, with a strong transition at sSFR$\sim10^{-9.5}$\,yr$^{-1}$. This is essentially the split between SFS galaxies with constant SFH, and `low mass passive' galaxies with constant SFH. The low mass passive galaxies formed the bulk of their stars much earlier in the Universe and have had largely constant, and low, SFRs for the last 7\,Gyr. As such, they will have occupied largely the same position in the sSFR-M$_{\star}$ plane for their entire history. By contrast, the higher sSFR population formed the bulk of their stars recently and therefore have had a significant increase in their SFR over the last 7\,Gyr. These are particularly transitory objects that will rapidly evolve through the plane, growing quickly in stellar mass \citep[$e.g.$ see][]{Caplar19}.  

What is potentially more interesting, is the consequence of the fact that all galaxies with constant  $\Delta$SFH$_{200\mathrm{\,Myr}}$ and low sSFR have had very little change in their SFRs over the last 7\,Gyr. This means that the large variation in sSFR seen at the low stellar mass end, is largely defined by some \textit{early} Universe condition of the galaxy, not something that has happened since. $i.e.$ their SFR was set in the early Universe and has stayed relatively the same since. What's more, the value of this sSFR does not seem well-correlated with stellar mass, as there is a huge diversity in sSFR values at a given stellar mass.  This of course assumes that the \textsc{ProSpect} analysis produces robust SFHs for these galaxies and the associated caveats with this. However, it is worth speculating on what this might be. Potentially their star formation is limited by their available gas supply, which is turn could be linked to halo/sub-halo mass, merger history or environmental history \citep[$e.g.$][]{Kawata08, Darvish16, Voort17}. This is just mere speculation, but this once again suggests that further study of the environment of these sources will likely yield insights into why they show a broad range of sSFRs at a fixed stellar mass - are these galaxies that had their star-formation suppressed at relatively early time (by environmental forces?) and have therefore not evolved significantly since that point?

\section{SFH stochasticity in the \textsc{shark} semi-analytic model}                                       
\label{sec:shark}

\begin{figure*}
\begin{center}
\includegraphics[scale=0.8]{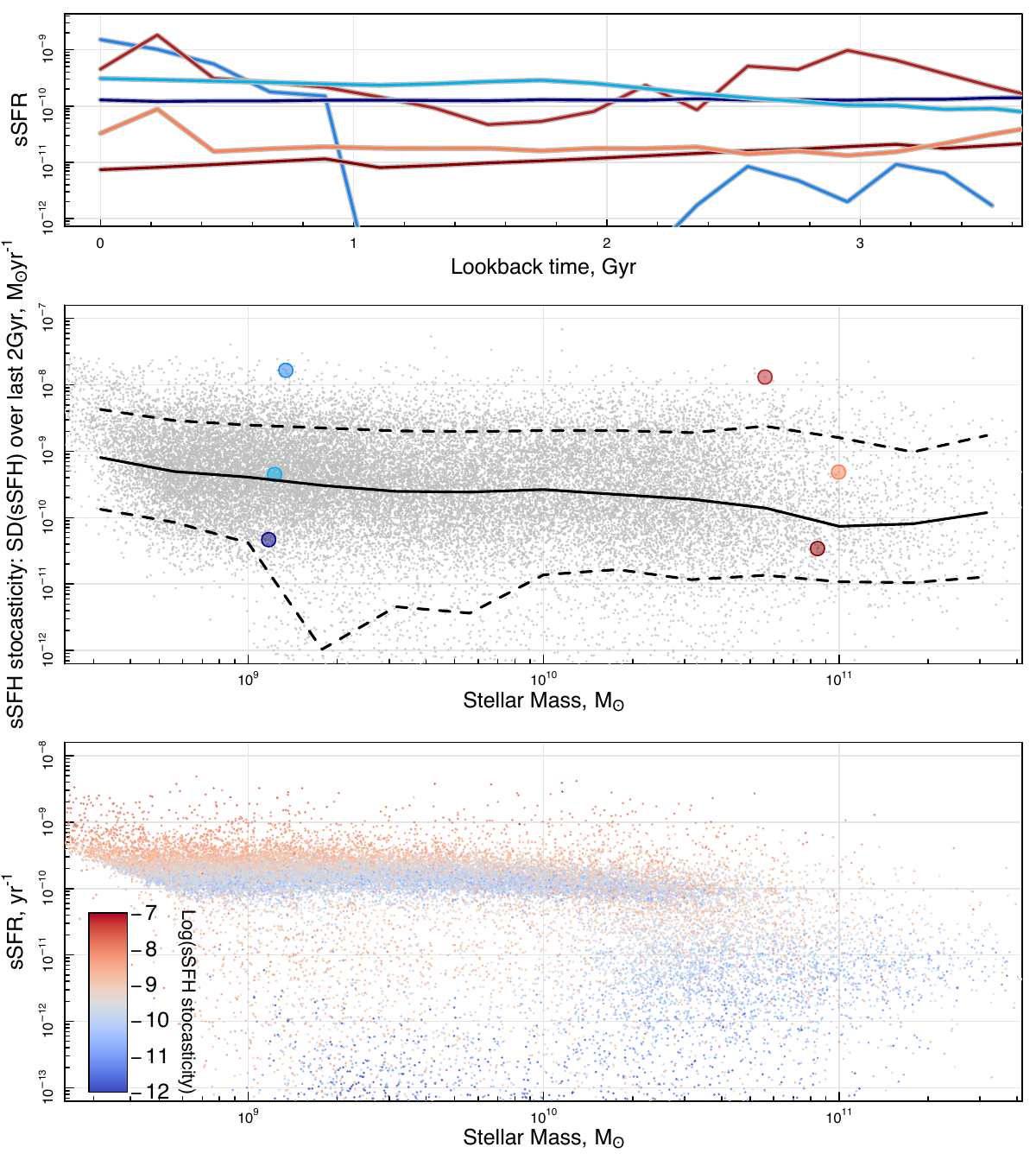}

\caption{sSFH stochasticity in the \textsc{shark} semi-analytic model. Middle panel: The y-axis shows the standard deviation in specific-SFH over the last 2\,Gyr for all z=0 \textsc{shark} galaxies (see text for details), with the running median and 0.1-0.9 range shown as solid and dashed lines respectively. We see only a weak trend of increasing stochasticity with lower stellar masses. The top panel shows selected sSFHs at low (blue) and high (red) stellar masses for a range of stochasticities. The position of these galaxies is displayed in the middle panel. Bottom panel: the sSFR-M$^{\star}$ plane for \textsc{shark} galaxies colour-coded by recent sSFH stochasticity.}

\label{fig:stoastic}
\end{center}
\end{figure*}

\textcolor{black}{In this final Appendix we explore the stochasticity in the recent SFH of galaxies in the \textsc{shark} semi-analytic model  \citep{Lagos18b, Lagos24}, with the aim of showing that at low stellar masses there is a broad range of stochasticities from highly stochastic bursting/quenching systems, to systems with constant SFHs. This potentially provides additional weight to the results discussed in Section \ref{sec:Starbursts}, that we find that the large scatter in sSFR at the low stellar mass end is potentially a combination of stochastic system and systems with long duration, but constant SFHs. First, we take all $z=0$ galaxies in \textsc{shark} and measure the recent stochasticity of their specific-SFH (sSFH) over the last 2\,Gyr. To do this we first extract a linear trend from the sSFH and then measure the standard deviation of the sSFH. The middle panel of Figure \ref{fig:stoastic} then shows this stochasticity as a function of stellar mass, with the running median (solid line) and 0.1-0.9 range (dashed lines). While there is a somewhat weak trend with stellar mass, we find that a broad range of stochasticities are found at all stellar masses. To highlight the variations in sSFH at a given stellar mass, the top panel shows a selection of galaxies at low stellar mass (blue) and high stellar mass (red). The positions of these galaxies are highlighted in the middle panel. These show that there are a broad range of SFHs at each stellar mass, ranging from constant, to highly stochastic. In the bottom panel, we then show the sSFR-M$^{\star}$ plane for these \textsc{shark} galaxies, colour-coded by recent stochasticity. This shows that while there is a similar distribution of stochasticities as a function of stellar mass, there are differences in where these galaxies lie in sSFR. Most notably the majority of low stellar mass galaxies with low stochasticity sit on the SFS, not below it - as we find in our observations.}

\bsp	% typesetting comment
\label{lastpage}
\end{document}